\documentclass[aps,prl,twocolumn,superscriptaddress]{revtex4-2}
\usepackage{graphicx,amsmath,amssymb,bm}

\usepackage{braket}
\usepackage{hyperref}
\usepackage{multirow} 

\DeclareMathOperator{\Tr}{Tr}

\usepackage{xcolor}
\definecolor{lightcoral}{rgb}{0.94117647, 0.50196078, 0.50196078}
\definecolor{cornflowerblue}{rgb}{0.39215686, 0.58431373, 0.92941176} 
\hypersetup{colorlinks=true, linkcolor=lightcoral, citecolor=cornflowerblue, urlcolor=cornflowerblue}

\begin{document}

\title{Fast Quantum Gates for Neutral Atoms Separated by a Few Tens of Micrometers}

\author{Matteo Bergonzoni}
\email{bergonzoni@unistra.fr}
\affiliation{University of Strasbourg and CNRS, CESQ and ISIS (UMR 7006), 67000 Strasbourg, France}

\author{Rosario Roberto Riso}
\affiliation{Department of Chemistry, Norwegian University of Science and Technology, 7491 Trondheim, Norway}
\affiliation{University of Strasbourg and CNRS, CESQ and ISIS (UMR 7006), 67000 Strasbourg, France}

\author{Guido Pupillo}
\email{pupillo@unistra.fr}
\affiliation{University of Strasbourg and CNRS, CESQ and ISIS (UMR 7006), 67000 Strasbourg, France}

\date{\today}

\begin{abstract}
We present a theoretical scheme for a family of fast and high-fidelity two-qubit iSWAP gates between neutral atoms separated by more than 20 $\mu$m, enabled by resonant dipole–dipole spin-exchange interactions between Rydberg states. The protocol harnesses coherent excitation–exchange–deexcitation dynamics between the qubit and the Rydberg states within a single and smooth laser pulse, in the presence of strong dipole-dipole interactions. We utilize optimal control methods to achieve theoretical gate fidelities and durations comparable to blockade-based gates in the presence of relevant noise, while extending the effective interaction range by an order of magnitude. This enables entanglement well beyond the blockade radius, offering a route toward fast, high-connectivity quantum processors.

\end{abstract}

\maketitle

Neutral atoms trapped in optical tweezers are a leading platform for quantum computing, enabling low-defect arrays of thousands of atoms with arbitrary geometries, long qubit coherence times, and fast, high-fidelity two-qubit gates \cite{Saffman_2010, Morgado_2021, Evered_2023, Ma_2023, Tsai_2025, Muniz_2025}. Most two-qubit gates rely on the Rydberg blockade, where van der Waals (vdW) interactions $V_{\rm{vdW}}(R)\sim R^{-6}$ between Rydberg-excited states suppress multiple excitations of nearby atoms \cite{Jaksch_2000, Jandura_2022, Levine_2019, Evered_2023, Ma_2023, Tsai_2025, Radnaev_2025, Muniz_2025,Peper_2025}, limiting the interaction range to a few micrometers. Alternative approaches such as the Rydberg antiblockade with vdW \cite{Wu_2021, Wu_2021_b, Li_2024} or dipole-dipole $J(R)\sim R^{-3}$ \cite{Su_2021} interactions have achieved similar ranges. Extending the entangling range without sacrificing speed or fidelity is however crucial for scalable, error-corrected architectures \cite{Morgado_2021,Saffman_2025}. Experimentally, entangling atoms over tens of micrometers is achieved via qubit shuttling — moving the tweezers — on hundred-microsecond timescales \cite{Saffman_2025, Pichard_2024, Zhou_2025}. Alternatively, DC or microwave fields can tune the Förster defect between selected Rydberg states enhancing the interaction strength, but such schemes are highly sensitive to field fluctuations and atomic positioning \cite{Ashkarin_2025, Kurdak_2025}. Another approach employs ancillary atoms to mediate interactions, at the cost of increased atomic overhead and longer gate times \cite{Cesa_2017, Sun_2024, Doultsinos_2025}.

\begin{figure}[t]
    \centering
    \includegraphics[width=0.97\columnwidth]{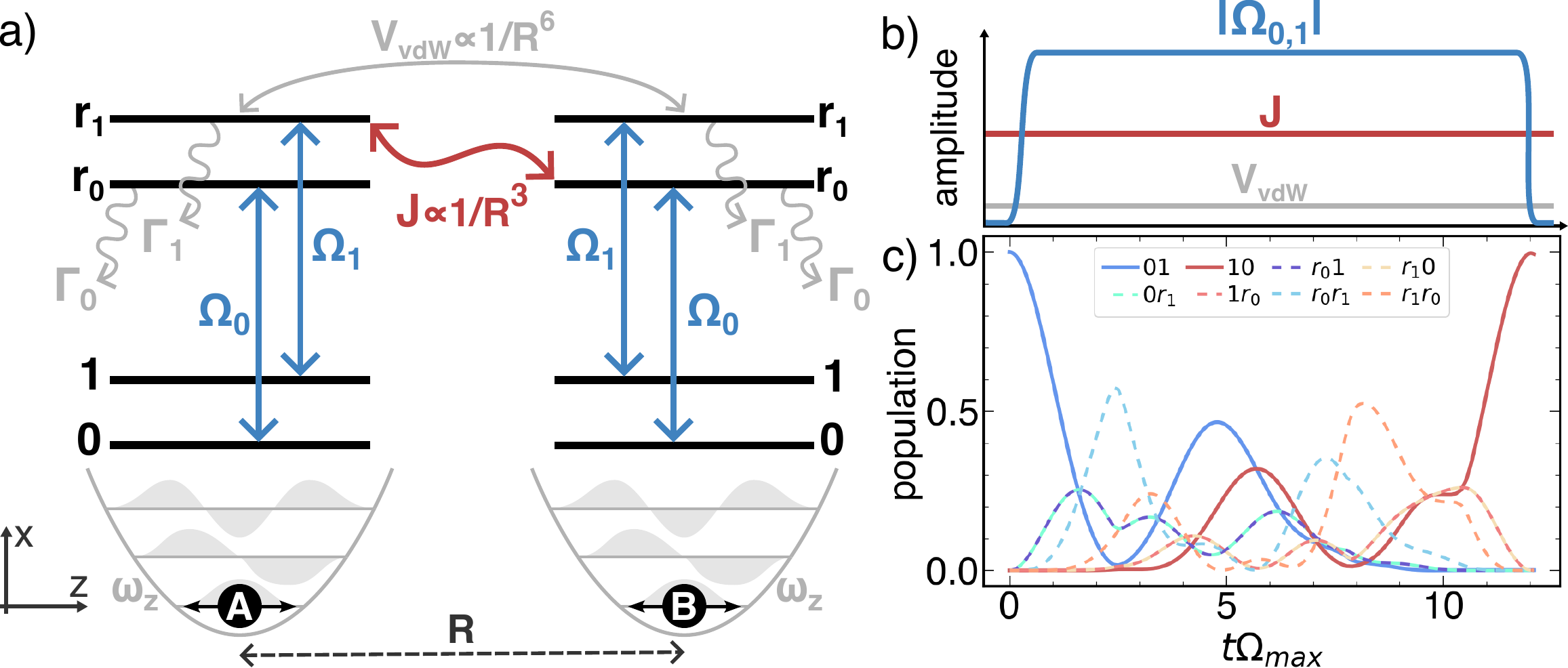}
    \caption{iSWAP gate with Rydberg atoms. \textbf{(a)} Two atoms separated by a distance $R$, irradiated by two laser fields $\Omega_{0,1}$, with a resonant dipolar coupling $J$. Rydberg decay $\Gamma_{0,1}$, atomic motion, and vdW interactions $V_{\text{vdW}}$ are also included. \textbf{(b)} A single optimal laser pulse $\Omega_{0,1}$ is applied, with dipolar and vdW interactions present from the outset. \textbf{(c)} Evolution of the two-atom state populations during the pulse for $\Omega_{\rm{max}}/J=2.1$ (Fig.~\hyperref[fig:pulses]{\ref{fig:pulses}(c)}), starting from the state $\ket{01}$.}
    \label{fig:level_scheme}
\end{figure}

Here, we introduce a family of fast two-qubit gates based on resonant dipolar exchange interactions $J(R)\sim R^{-3}$ between atoms separated by tens of micrometers \cite{Reinhard_2007, Chew_2022, Mhaignerie_2025, Emperauger_2025, Giudici_2025, Su_2021, Young_2021, Crane_2021}. The protocol implements coherent excitation to, and de-excitation from, the Rydberg manifold — where rapid resonant exchange occurs — from long-lived ground-state qubits, enabling sub-microsecond iSWAP gates (Fig.~\ref{fig:level_scheme}).
While reminiscent of Rydberg antiblockade schemes \cite{Li_2024, Wu_2021, Wu_2021_b, Su_2021}, our approach introduces qualitatively new elements: a single global, continuous laser pulse with a {\it time-varying phase} optimized via quantum optimal control to realize gates that are {\it time-optimal} and {\it robust} against dominant error sources, including vdW interactions, spontaneous emission. The resulting fidelities and gate times are comparable to blockade-based protocols, while extending the interaction range by nearly an order of magnitude, enabling long-range entanglement in near-term quantum memories and processors. Similar schemes for iSWAP gates with Sr atoms have also been proposed in Ref.~\cite{Ildefonso_2025}.

\begin{figure*}[t]
    \includegraphics[width=0.93\textwidth]{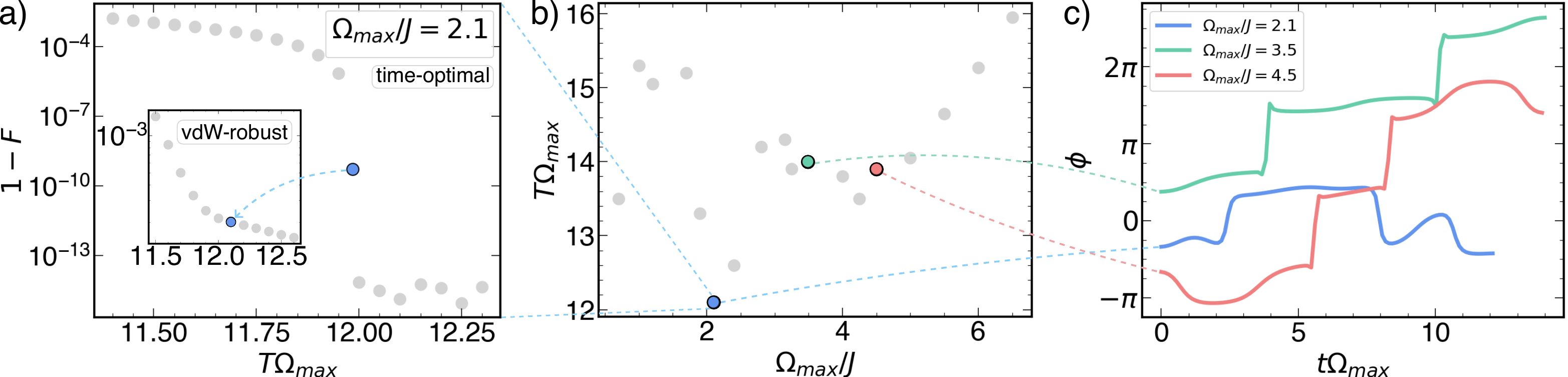}
    \caption{Time-optimal and vdW-robust pulses. \textbf{(a)} Minimum infidelity $1-F$ found by GRAPE for pulses of duration $T$, in units of the Rabi frequency $\Omega_{\rm max}$, for the ratio $\Omega_{\rm{max}}/J=2.1$. The time-optimal pulse has duration $T\Omega_{\rm{max}}=11.95$, while trying to make it robust to vdW interactions the pulse durations is increased to $T\Omega_{\rm{max}}=12.1$ (inset). \textbf{(b)} Optimal duration $T$, in units of $\Omega_{\rm max}$ versus coupling ratio $\Omega_{\max}/J$ of the vdW-robust pulses. The range $R$ and fidelity $F$ obtained for a realistic choice of experimental parameters are shown in Fig.~\ref{fig:fidelity}. \textbf{(c)} Optimal phase profile $\phi(t)$ of three vdW-robust pulses.}
    \label{fig:pulses}
\end{figure*}
We consider two atoms, labeled $\alpha=A$ and $B$, separated by distance $R$ (Fig.~\hyperref[fig:level_scheme]{\ref{fig:level_scheme}(a)}). Logical states $\ket{0}_{\alpha}$ and $\ket{1}_{\alpha}$ are encoded in long-lived ground or metastable states of the atoms. The atoms are simultaneously irradiated by two global fields with complex Rabi frequency $|\Omega_j(t)|e^{i\phi_j(t)}$, driving the transition $\ket{j}_{\alpha}\leftrightarrow\ket{r_j}_{\alpha}$ ($j=0,1$), with $\ket{r_0}_{\alpha}$ and $\ket{r_1}_{\alpha}$ two Rydberg states. A detailed discussion on the implementation of the $\ket{j}_\alpha\leftrightarrow\ket{r_j}_\alpha$ transition, including multi-photon processes, is presented in the Suppl. Mat. \cite{SM}. The Rydberg states of the two atoms are coupled by the dipolar spin-exchange interaction $J$, resulting in the following Hamiltonian in the rotating frame
\begin{align}
    \nonumber
    H(t) = & \sum_{j,\alpha} \frac{|\Omega_j(t)|}{2}\left( e^{i\phi_j(t)} \ket{j}\bra{r_j}_\alpha + \text{h.c.}\right)\\
    \label{H}
    & + J\left( \ket{r_0r_1}\bra{r_1r_0} + \text{h.c.}\right),
\end{align}
with $\ket{r_0 r_1} \equiv \ket{r_0}_A \otimes \ket{r_1}_B$ and $\hbar =1$. Equation~\eqref{H} neglects spontaneous emission, atomic motion, photon recoil, and residual vdW interactions; their impact and mitigation are discussed below. In principle, Eq.~\eqref{H} enables an iSWAP gate,
$U_{\mathrm{iSWAP}}=\ket{00}\bra{00}+\ket{11}\bra{11}+i(\ket{01}\bra{10}+\ket{10}\bra{01})$,
which, together with single-qubit rotations, forms a universal gate set. The simplest implementation, analogous to schemes with polar molecules \cite{Ni_2018, Holland_2023, Bao_2023, Picard_2025, Ruttley_2025, Bergonzoni_2025}, uses a $\pi$ pulse to excite the atoms to Rydberg states, free evolution under $J$ to generate entanglement, and a second $\pi$ pulse for de-excitation (see Suppl. Mat. \cite{SM}). However, in Rydberg systems the dipolar interaction is typically comparable to or larger than the Rabi frequency ($J \gtrsim |\Omega_{0,1}|$), perturbing the $\pi$-pulse dynamics. Thus, we propose a protocol where a single continuous laser pulse whose amplitude and — crucially — phase are time-dependent and designed via quantum optimal control methods to be time-optimal and robust with respect to these interactions during excitation and de-excitation, as well as vdW interactions in the excited states and Rydberg-state spontaneous emission. This enables high-fidelity gates compatible with most quantum error correction schemes over tens of micrometers in the presence of all relevant noise sources, including laser photon recoil, atomic motion and scattering from intermediate states in multi-photon transitions. Similarly to Rydberg blockade gates \cite{Jandura_2022,Bluvstein_2023}, the continuous global pulse irradiating both atoms is expected to simplify experiments.

As an example, we consider $^{87}$Rb, where qubit states can be encoded in the hyperfine levels $\ket{0}_\alpha \equiv \ket{F=1,m_F=0}_\alpha$ and $\ket{1}_\alpha \equiv \ket{F=2,m_F=0}_\alpha$ of the  ground state manifold $5S_{1/2}$, with $F$  the total angular momentum and $m_F$ its projection along the quantization (interatomic) axis $z$. The Rydberg states $n\rm{S}_{1/2}$ with $m_j=1/2$ and $n\rm{P}_{3/2}$ with $m_j=3/2$ are chosen as $\ket{r_0}_\alpha$ and $\ket{r_1}_\alpha$, respectively, with $n$ the principal quantum number. 
This results in dipolar interaction energies $J$ in the range $0.1 \lesssim J/2\pi \lesssim 1\,\rm{GHz}$, for the considered quantum numbers $n \gtrsim 50$ and interatomic distances $R\sim 3\text{–}30\,\mathrm{\mu m}$. We consider realistic Rabi frequencies $1 \leq|\Omega_{0,1}|/2\pi \leq 10\,\mathrm{MHz}$.
\begin{table*}[t]
\centering
\renewcommand{\arraystretch}{1.2}
\begin{tabular}{c|c || c | c | c | c | c | c | c | c||c}
\hline\hline
\multicolumn{2}{c||}{\raisebox{-1.2ex}[0pt][0pt]{Noise term}} & Atomic & vdW & Decay & $z$-motion & \multicolumn{2}{c|}{Recoil (detuning)} & \multicolumn{2}{c||}{Recoil (coupling)} & All \\
\cline{7-10}
\multicolumn{2}{c||}{} & transitions & $H_{\rm vdW}$ & $H_\Gamma$ & $H_{J}^{(1,z)}+H_{\rm vdW}^{(1,z)}$ & $\Delta_j=0$ & $\Delta_j=-\frac{2\pi^2}{m\lambda_j^2}$ & $\bar{\ell}=x$ & $\bar{\ell}=z$ &  \\
\hline
\multirow{2}{*}{$1-F$} 
& TO & $5.5\times10^{-3}$ & $1.3\times10^{-2}$ & $7.0\times10^{-4}$ & $4.3\times10^{-5}$ & $1.3\times10^{-5}$ & $<10^{-7}$ & $2.3\times10^{-5}$ & $4.9\times10^{-5}$ & $1.9\times10^{-2}$\\
\cline{2-11}
& vdW-R & $4.1\times10^{-3}$ & $2.0\times10^{-4}$ & $7.0\times10^{-4}$ & $3.7\times10^{-5}$ & $8.7\times10^{-6}$ & $<10^{-7}$ & $1.9\times10^{-5}$ & $5.1\times10^{-5}$ & $5.1\times10^{-3}$ \\
\hline\hline
\end{tabular}
\caption{Error budget of the time-optimal (TO) and van der Waals–robust (vdW-R) pulses for $\Omega_{\rm max}/J = 2.1$.
Reported are infidelities $1-F$ in the presence of individual perturbations: realistic atomic transitions (off-resonant couplings and intermediate state scattering), vdW interactions, Rydberg decay, atomic motion along the interatomic $z$ axis, and photon recoil — considering both its effects: detuning (compensated and uncompensated) and coupling between internal and motional states (for laser orientations along $x$ and $z$). Experimental parameters used in the simulations: Rabi frequency $\Omega_{\rm max}/2\pi=10\,\mathrm{MHz}$, principal quantum number $n=100$ of $^{87}\rm{Rb}$, trapping frequencies $\omega_z/2\pi = 100\,\mathrm{kHz}$, $\omega_x = \omega_z/5$ and temperature $T_{\rm temp}=1\,\mathrm{\mu K}$.
}
\label{tab:error_budg}
\end{table*}

\paragraph{Time-optimal gate.---}
As a first step, we demonstrate a time-optimal scheme for implementing $U_{\mathrm{iSWAP}}$ with Eq.~\eqref{H}, i.e., including only the combined effects of $J$ and the laser drive. Time optimality minimizes spontaneous emission, a fundamental error source for Rydberg atoms \cite{Jandura_2022, Evered_2023, Ma_2023, Tsai_2025, Peper_2025, Muniz_2025}. As in Rydberg blockade schemes \cite{Jandura_2022, Pagano_2022, Mohan_2023}, we define a cost function $1-F$, with $F$ the Bell state fidelity
\begin{equation}
    \label{F}
    F = \frac{1}{16}\left| \sum_{q} \braket{\psi_q| R^{\otimes2}U_{\rm{iSWAP}}|q} \right|^2,
\end{equation}
where $\ket{\psi_q}$ is the actual two-qubit final state obtained by applying the pulse to $\ket{q}$, from the computational basis $\ket{q} \in \{\ket{00}, \ket{01}, \ket{10}, \ket{11}\}$. $\ket{\psi_q}$ is therefore compared to the target state $U_{\rm{iSWAP}}\ket{q}$ up to a final global single-qubit operation $R(\theta,\varphi,\lambda) = e^{i(\varphi+\lambda)/2} R_z(\varphi) R_y(\theta) R_z(\lambda)$, where $\varphi$, $\theta$, and $\lambda$ are rotation angles treated as additional optimization parameters. The cost function is numerically minimized using quantum optimal control methods — here we use the GRAPE algorithm \cite{Khaneja_2005, Jandura_2022}. The time-optimal pulse is defined as the shortest pulse that achieves zero infidelity $1-F$, within numerical precision, once a specific Hamiltonian — encoding all the relevant physical interactions and control fields — is fixed. We apply this approach to a wide range of experimentally relevant ratios $0\lesssim|\Omega_{0,1}|/J\lesssim 10$. 

\begin{figure*}[t]
    \includegraphics[width=0.98\textwidth]{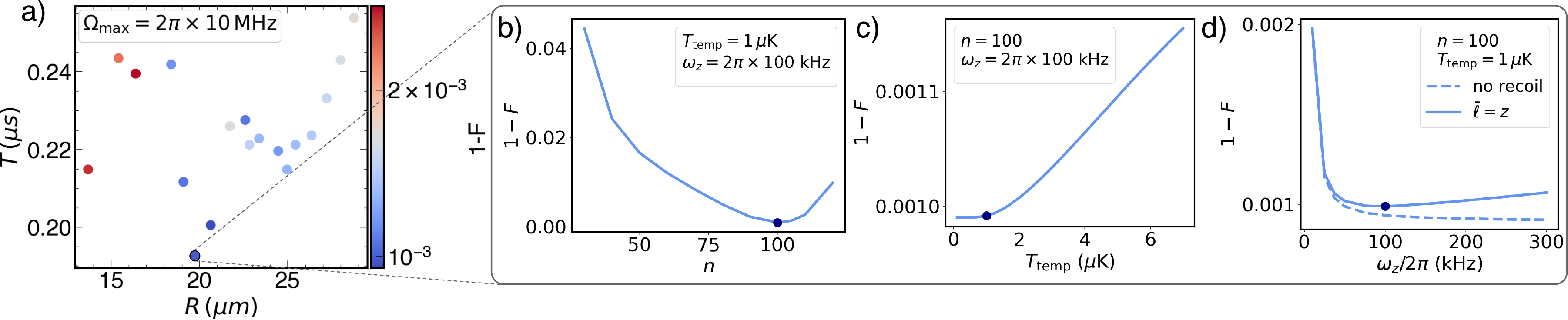}
    \caption{Infidelity of the vdW-robust pulses for a fixed Rabi frequency $\Omega_{\rm max}/2\pi=10\,\mathrm{MHz}$ — without accounting for the errors from realistic atomic transitions. \textbf{(a)} Heat map of gate infidelity $1-F$ (color scale on the right) for several coupling ratios $\Omega_{\max}/J$, with $1.5 \leq J \leq 15\,\mathrm{MHz}$. \textbf{(b)-(d)} Fidelity of the vdW-robust pulse for $\Omega_{\max}/J=2.1$ as a function of (b) the principal quantum number $n$, (c) the atomic temperature $T_{\rm temp}$, and (d) the radial trapping frequency $\omega_z/2\pi$, in absence of photon recoil and with the laser orientated along the interatomic direction ($\bar{\ell}=z$).}
    \label{fig:fidelity}
\end{figure*}

Our first key result is that, for $0\lesssim|\Omega_{0,1}|/J \lesssim 10$ time-optimal pulses are those for which the amplitudes of the two Rabi frequencies are equal to the maximum available Rabi frequency in experiments, which here we assume identical for the two lasers, i.e., $|\Omega_0(t)|=|\Omega_1(t)|=\Omega_{\rm{max}}$. In addition, we find that time optimality also requires same lasers phases, i.e., $\phi_0(t)=\phi_1(t)=\phi(t)$, with  $\phi(t)$ the key variable to be varied in time to minimize $1-F$, which has not been done before.
Figure~\hyperref[fig:fidelity]{\ref{fig:pulses}(a)} shows an example result of the infidelity $1-F$ versus the total gate duration $T$ for a given ratio $\Omega_{\rm{max}}/J=2.1$, with $T$ in units of $\Omega_{\rm{max}}$. The figure shows that the infidelity decreases abruptly to numerical zero around $T\Omega_{\rm{max}}=11.95$ (blue dot), corresponding to the time-optimal pulse. Varying the ratio $\Omega_{\rm max}/J$ results in different optimal phase profiles and gate durations, with the shortest pulse obtained for $\Omega_{\rm max}/J=0.7$, corresponding to $T\Omega_{\rm max}=8.2$. This is essentially the same timescale of the time-optimal blockade gate ($T\Omega_{\rm{max}}=7.5$) \cite{Jandura_2022} or the recently proposed short-range resonant dipole-dipole CZ gate ($T\Omega_{\rm{max}}=5.95$) \cite{Giudici_2025}. However, here the range of the gate scales as $R\sim\sqrt[3]{n^4(\Omega_{\rm{max}}/J)/\Omega_{\rm{max}}}$, which allows to reach distances of tens of micrometers. For example, for $\Omega_{\rm{max}}/2\pi=10\,(5)\,\mathrm{MHz}$, $\Omega_{\rm{max}}/J=2.1\,(4.25)$ and $n=100$, we obtain $R=20\,(31)\,\mathrm{\mu m}$. 
We find that the optimal $T$ versus $\Omega_{\rm{max}}/J$ is not monotonic (see Suppl. Mat. \cite{SM}), but generally increases with $\Omega_{\rm{max}}/J$. While shorter pulses would minimize spontaneous emission, they are more susceptible to atomic motion and vdW interactions, and therefore do not necessarily yield the optimal solution in realistic scenarios, which we address next.

\paragraph{Imperfections and robustness.---}
The relevant Hamiltonian to realize $U_{\mathrm{iSWAP}}$ gates with neutral atoms beyond Eq.~\eqref{H} reads $ H' = H_\Omega' + H_J'+H_{\rm vdW}' + H_K + H_\Gamma$, with
\begin{align}
   \label{H_recoil}
    & H_\Omega' =  \sum_{j,\alpha}\frac{\Omega_j}{2} \left[e^{i\phi_j+i\eta_j(a_{\bar{\ell},\alpha}+{a_{\bar{\ell},\alpha}}^\dagger)}\ket{j}\bra{r_j}_\alpha+\text{h.c.}\right]\nonumber\\
   & H_J' =  J \left[1-3\frac{z_A-z_B}{R}\right] \left(\ket{r_0r_1}\bra{r_1r_0}+\text{h.c.}\right)\nonumber\\
  &  H_{\rm vdW}' =  \sum_{i,j} V_{ij} \left[1-6\frac{z_A-z_B}{R}\right] \ket{r_i r_j} \bra{r_i r_j}\nonumber\\
 &    H_K = - \sum_{\alpha, \ell} \frac{p^2_{\alpha,\ell}}{2m},  \;\; H_\Gamma =  -\frac{i}{2} \sum_{j,\alpha} \Gamma_j \ket{r_j}\bra{r_j}_\alpha
.
\end{align}
The term $H_\Omega'$ in Eq.~\eqref{H_recoil} 
includes the effect of photon recoil in the laser driving Hamiltonian, characterized by the Lamb–Dicke parameter $\eta_j$, and with ${a_{\bar{\ell},\alpha}}^\dagger$ and $a_{\bar{\ell},\alpha}$ the ladder operators for the harmonic trap of atom $\alpha$ along the laser propagation direction $\bar{\ell}$ \cite{Robicheaux_2021, Giudici_2025}. 
The term $H_J'$ includes the first-order correction in the dipole-dipole interaction $J$ of Eq. \eqref{H}, arising from atomic motion along the interatomic direction $z$, where $z_\alpha$ is the position operator of atom $\alpha$ in its radial harmonic trap (see Suppl. Mat. \cite{SM}). $H_{\rm vdW}'$ describes  off-resonant van der Waals interactions and their first-order position-dependent corrections $H_{\rm vdW}^{(1,z)}$ in the $z$ direction.
$H_K$ is the kinetic operator, with $m$ the atomic mass and $p_{\alpha,\ell}$ the momentum operator of atom $\alpha$ along the three axis $\ell = x, y, z$. The trapping potential is here omitted as optical tweezers are usually turned off during the Rydberg excitation pulses \cite{Zuo_2009, Evered_2023}. Finally, $H_\Gamma$ in Eq.~\eqref{H_recoil} is a non-Hermitian term describing decay from Rydberg states with rates $\Gamma_{0,1}$. 

Photon recoil has a twofold effect: First, it introduces a constant detuning $\Delta_j = 2\pi^2/(m\lambda_j^2)$ corresponding to the  kinetic energy imparted by a photon of laser $j$, with wavelength $\lambda_j$. Second, it induces a coupling between motional degrees of freedom along the $\bar{\ell}$ direction and the internal states of the atom, with strength $\sim \sqrt{\omega_{\bar{\ell}}/(m\lambda_j^2)}$ and with $\omega_{\bar{\ell}}$ the trapping frequency in the laser direction (see Suppl. Mat. \cite{SM}). This results in  decoherence of the internal states by  entanglement of motional and electronic states.
$H_J'$ results in further coupling between motional degrees of freedom and internal states of the atom — with coupling strength $\sim J/(R\sqrt{m\omega_z})$ — due to quantum mechanical motional spread of the wavefunctions.
Similar effects are observed for van der Waals interactions $H_{\rm vdW}$.
Table~\ref{tab:error_budg} provides an example of error budget resulting from the terms in Eq. \eqref{H_recoil} for the time-optimal (TO) pulse with $\Omega_{\rm{max}}/J = 2.1$, for a choice of realistic experimental parameters. The table shows that vdW interactions largely dominate the error budget, accounting for an infidelity $1-F \sim 10^{-2}$, while all other error sources provide orders of magnitude smaller contributions (see Suppl. Mat. \cite{SM} for details). 

In this work, we therefore proceed to obtain higher-fidelity gates by making the TO pulses robust against vdW interactions. This is achieved by including the latter directly in the optimization, using the TO pulses as initial guess \cite{Giudici_2025}.
In general, for small ratios $\Omega_{\rm max}/J \lesssim 1$ — corresponding to smaller interatomic distances $R$ — the numerical optimization fails to find a robust pulse with a duration $T$ comparable to the TO one. For larger ratios $\Omega_{\rm{max}}/J \gtrsim 3$, however, we find pulses with high fidelity while maintaining almost the same pulse duration, as a result of the short-range nature of vdW interactions. Figure~\hyperref[fig:pulses]{\ref{fig:pulses}(b)} shows the minimum vdW-robust gate duration $T$ as a function of $\Omega_{\rm{max}}/J$: For $\Omega_{\rm{max}}/J = 2.1$, the robust pulse duration is $T\Omega_{\rm{max}} = 12.1$. This is less than 2\% longer than the corresponding TO pulse (see Fig.~\hyperref[fig:pulses]{\ref{fig:pulses}(a)}), but with infidelity well below $10^{-3}$ in the presence of vdW interactions only. Increasing the duration $T$ of the robust pulse can further decrease the infidelity (see inset in Fig.~\hyperref[fig:pulses]{\ref{fig:pulses}(a)}).   Figure~\hyperref[fig:pulses]{\ref{fig:pulses}(c)} plots the optimal phase $\phi(t)$ for three representative pulses: In all cases, the pulses are smooth in time, which should make them easier to implement in experiments. Jumps of $\pi$ in the phase for $\Omega_{\rm max}/J>2.1$ can in principle be removed by re-optimizing the pulse for the detuning — instead of the phase only — and fixing a cutoff, with minimal loss of pulse performance \cite{Pecorari_2025_3qubit}.

Figure~\hyperref[fig:fidelity]{\ref{fig:fidelity}(a)} analyzes of a set of vdW–robust pulses for different values of the ratio $\Omega_{\rm{max}}/J$, including all terms in Eq.~\eqref{H_recoil}. The pulse with $\Omega_{\rm{max}}/J = 2.1$ — assuming a Rabi frequency $\Omega_{\rm{max}}/2\pi = 10\,(5)\,\mathrm{MHz}$, radial trapping frequency $\omega_z/2\pi = 100\,\mathrm{kHz}$, temperature $T_{\rm{temp}} = 1\,\mathrm{\mu K}$, and principal quantum number $n = 100$ of $^{87}\rm{Rb}$ — yields an infidelity $1-F < 1\,(3)\times10^{-3}$, at an interatomic distance $R = 20\,(31)\,\mathrm{\mu m}$ with total pulse duration $T = 197\,(430)\,\mathrm{ns}$. This is comparable to that of CZ gates based on the blockade mechanism \cite{Jandura_2022, Ma_2023, Evered_2023}. To cover the same distance the state-of-the-art qubit transport would require times that are three orders of magnitude longer \cite{Saffman_2025}. The fidelity in Fig.~\hyperref[fig:fidelity]{\ref{fig:fidelity}(a)} remains high, $1-F \leq 2\times10^{-3}$, even for larger ratios $\Omega_{\rm{max}}/J \lesssim 7$, allowing the implementation of entangling pulses between even more distant atoms, $R \approx 30\,\mathrm{\mu m}$, on sub-$\mu$s timescales.

In Figs.~\hyperref[fig:fidelity]{\ref{fig:fidelity}(b)-(d)}, the fidelity of the robust pulse $\Omega_{\rm{max}}/J=2.1$ is analyzed as a function of the principal quantum number $n$, the temperature $T_{\rm{temp}}$, and the radial trapping frequency $\omega_z$. Increasing $n$ shortens the pulse duration, since $J\sim n^4$, and increases the Rydberg lifetime $\Gamma\sim n^{-3}$, thus reducing infidelity, provided the pulse remains robust against the corresponding vdW interactions; otherwise, larger $n$ amplifies their perturbative effect $V_{\rm{vdW}}\sim n^{11}$ (see Fig.~\hyperref[fig:fidelity]{\ref{fig:fidelity}(b)}). This improvement is partially offset by the decrease of the Rabi frequency $\Omega_{\rm{max}}\sim n^{-3/2}$ \cite{Saffman_2010}. Lower temperatures reduce the occupation of excited motional states and hence position uncertainty (see Fig.~\hyperref[fig:fidelity]{\ref{fig:fidelity}(c)}). The trapping frequency also plays a crucial role: the displacement operator scales as $1/\sqrt{\omega_z}$, whereas the momentum kick grows as $\sqrt{\omega_{\bar{\ell}}}$. If lasers are aligned along the interatomic direction $\bar{\ell}=z$, a trade-off between position fluctuations and photon recoil leads to an optimal confinement near $\omega_z^*\approx90\,\text{kHz}$ (solid line in Fig.~\hyperref[fig:fidelity]{\ref{fig:fidelity}(d)}), as documented in Refs.~\cite{Robicheaux_2021, Pagano_2022, Giudici_2025, Emperauger_2025}. If the lasers are along the more weakly confined axial direction $\bar{\ell} = x$, with typically $\omega_z \approx 5 \omega_x$, photon recoil effects are suppressed and the optimal confinement becomes much larger (dashed line in Fig.~\hyperref[fig:fidelity]{\ref{fig:fidelity}(d)}).

In addition to the error sources in Eq.~\eqref{H_recoil}, one must account for errors from realistic atomic transitions  (see Suppl. Mat. \cite{SM}), in particular, scattering from the intermediate state $\ket{e}=\ket{6P_{3/2}, F=2,m_F=1}$ in the two-photon excitation to $\ket{r_0}$, as well as off-resonant couplings to other Rydberg states, such as $\ket{r_0'}=\ket{98D_{5/2},m_j=-3/2}$ and $\ket{r_1'}=\ket{100P_{3/2},m_j=1/2}$, that induce unwanted Stark shifts and population leakage. These effects increase the infidelity up to $1 - F \approx 5.1\,(8.4) \times 10^{-3}$ for $\Omega_{\rm max}/J=2.1\,(4.25)$ and $\Omega_{\rm max}=10\,(5)\,\mathrm{MHz}$, with an intermediate detuning $\Delta_e=40\,\Omega_{\rm max}$ and a magnetic field $B=50\,(20)\,\mathrm{G}$. The resulting fidelities remain compatible with most quantum error-correction schemes.

\paragraph{Conclusion.---}
We have demonstrated a protocol for time-optimal, high-fidelity iSWAP gates between neutral atoms separated by tens of micrometers. The scheme combines resonant dipolar spin-exchange interactions with optimally shaped laser pulses, achieving gate speeds comparable to the fastest CZ gates at few-$\mathrm{\mu m}$ distances. The fidelities reached are compatible with error-correction thresholds \cite{Fowler_2012}, with further improvements possible by mitigating residual errors listed in Table~\ref{tab:error_budg}. In particular, recoil-induced detuning can be corrected via laser frequency fine-tuning, and motional–electronic coupling reduced by aligning beams along the weakly confined axial direction. Using heavier species, such as $^{133}$Cs, further suppresses motional errors.
These results open several directions. The approach can be extended to $k$-qubit gates ($k>2$) via resonant dipole–dipole interactions, enhancing stabilizer efficiency \cite{Old_2025, Pecorari_2025} while reducing gate count and circuit depth \cite{Baker_2021}. Fast, long-range gates enable efficient quantum information transport across neutral-atom registers without intermediate operations, ancillary qubits, or atomic motion — crucial for modular quantum computing and error correction. Moreover, these gates can facilitate recently developed low-density parity-check (LDPC) codes, which require stabilizer qubits separated by large distances \cite{Tremblay_2022, Bravyi_2024, Xu_2024, Pecorari_2025, Poole_2025}. Exploiting long-range interactions, such codes achieve higher encoding rates, reduced qubit overhead, and larger code distances $d$, yielding lower logical error probabilities than surface codes of comparable size.

\vspace{3mm}
\noindent\textit{Acknowledgments.---} We are grateful to S. Whitlock for insightful discussions.
This research has received funding from the European Union’s Horizon Europe programme HORIZON-CL4-2021-DIGITAL-EMERGING-01-30 via the project 101070144 (EuRyQa) and from the French National Research Agency under the Investments of the Future Program projects ANR-21-ESRE-0032 (aQCess), ANR-22-CE47-0013-02 (CLIMAQS), and ANR-22-CMAS-0001 France 2030 (QuanTEdu-France).

\vspace{1mm}
\noindent See Supplemental Material \cite{SM} for details on the optimization algorithm, physical implementation, and fidelity.

\bibliography{main}

\nocite{Uhlmann_1976, Sibalic_2017, Chomaz_2023, Wall_2015, SciPy_2020, deKeijzer_2023, Nielsen_Chuang_2010, Boller_1991, Moller_2008, Hankin_2014, Li_2019, Saffman_2016, Jau_2016,  data_2025}

\onecolumngrid
\appendix
\clearpage

\renewcommand{\thesection}{S\arabic{section}}
\renewcommand{\thefigure}{S\arabic{figure}}
\renewcommand{\thetable}{S\arabic{table}}
\renewcommand{\theequation}{S\arabic{equation}}

\setcounter{section}{0}
\setcounter{figure}{0}
\setcounter{table}{0}
\setcounter{equation}{0}

\makeatletter
\renewcommand{\theHsection}{S\arabic{section}}
\renewcommand{\theHfigure}{S\arabic{figure}}
\renewcommand{\theHtable}{S\arabic{table}}
\renewcommand{\theHequation}{S\arabic{equation}}
\makeatother

\title{Supplemental Material for ``Fast Quantum Gates for Neutral Atoms Separated by a Few Tens of Micrometers''}

\begin{center}
\textbf{\large Supplemental Material for\\
``Fast Quantum Gates for Neutral Atoms Separated by a Few Tens of Micrometers''}
\end{center}

\section{Atomic transitions}

\subsection{\label{sec:choice_state} Main states}

In this work, we consider atoms of rubidium-87. In the ground state $5S_{1/2}$, the logical qubit states are encoded in the long-lived hyperfine levels $\ket{0} \equiv \ket{5S_{1/2}, F=1, m_F=0}$ and $\ket{1} \equiv \ket{5S_{1/2}, F=2, m_F=0}$, where $F$ denotes the total atomic angular momentum and $m_F$ its projection along the quantization axis defined by the applied magnetic field. The Rydberg states $nS_{1/2}$ with $m_j = 1/2$ and $nP_{3/2}$ with $m_j = 3/2$ are chosen as $\ket{r_0} = \ket{nS_{1/2}, m_j = 1/2}$ and $\ket{r_1} = \ket{nP_{3/2}, m_j = 3/2}$, respectively, with $n \gtrsim 50$ the principal quantum number.

The choice of $n$ plays a crucial role in determining both the duration and fidelity of the protocol, as it sets the scaling of the relevant physical quantities: the Rabi frequency $\Omega_{\mathrm{max}} \sim n^{-3/2}$, the decay rate $\Gamma \sim n^{-3}$, and the dipole–dipole and van der Waals (vdW) interaction strengths, which scale as $J \sim n^4$ and $V_{\mathrm{vdW}} \sim n^{11}$, respectively \cite{S_Saffman_2010}. Throughout most of this work, we adopt a high principal quantum number, $n = 100$. Tables~\ref{tab:atomic_data1} and~\ref{tab:atomic_data2} provide representative values of the relevant atomic parameters considered here, including the resonant dipole–dipole and vdW interaction strengths $J$ and $V_{\mathrm{vdW}}$, the corresponding dispersion coefficients $C_3$ and $C_6$, and the decay rates $\Gamma$.

Two specific implementations of our scheme, including the choice of intermediate states and laser polarizations, are detailed in the following section. We also note that, after the submission of this manuscript, a related work appeared in the literature \cite{S_Ildefonso_2025}, in which a similar protocol is implemented for strontium-88 atoms using metastable states. A comparable approach can also be developed for ytterbium-171 atoms \cite{S_Ma_2023}.
\begin{figure}[b]
    \centering
    \includegraphics[width=0.75\linewidth]{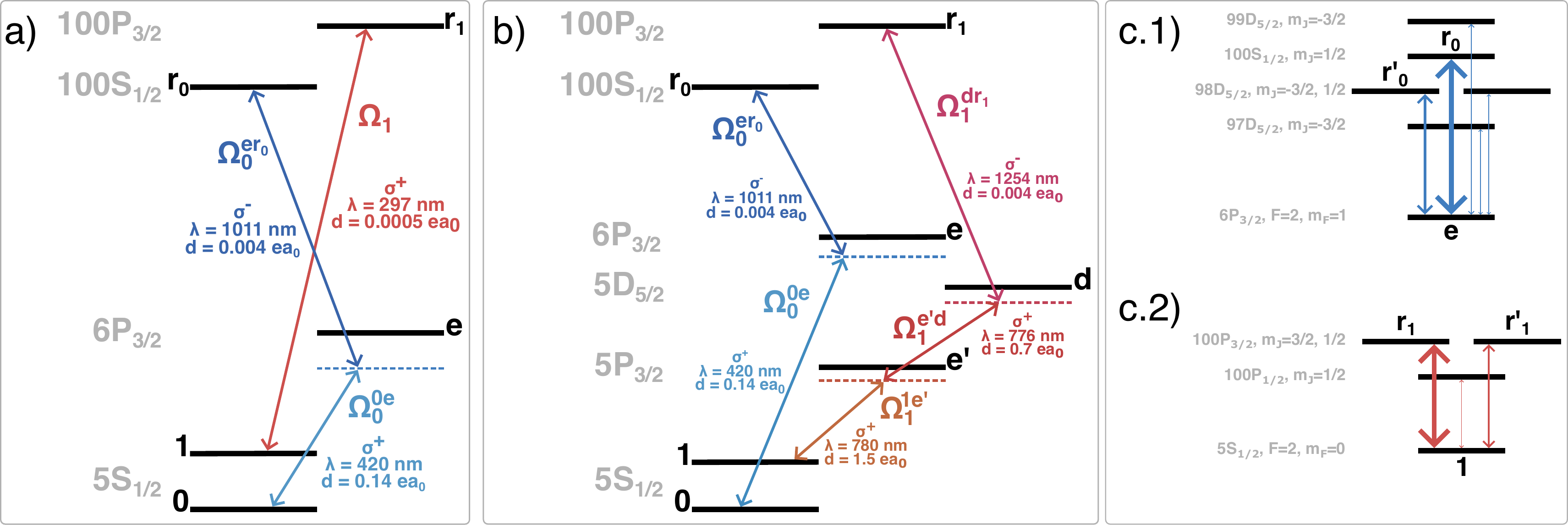}
    \caption{
    Atomic transitions used to realize the couplings $\ket{0}\leftrightarrow\ket{r_0}$ and $\ket{1}\leftrightarrow\ket{r_1}$. For each transition $\ket{a}\leftrightarrow\ket{b}$, we report the polarization $q=\pi,\sigma^{\pm}$, the wavelength $\lambda$, and the dipole matrix element $d = |\langle a | d_q | b \rangle|$. \textbf{(a)} A two-photon excitation implements the $\ket{0}\leftrightarrow\ket{r_0}$ transition via the intermediate state $\ket{e}$, while a single-photon UV excitation implements the $\ket{1}\leftrightarrow\ket{r_1}$ transition. \textbf{(b)} A two-photon excitation implements the $\ket{0}\leftrightarrow\ket{r_0}$ transition via the intermediate state $\ket{e}$, while a three-photon excitation implements the $\ket{1}\leftrightarrow\ket{r_1}$ transition via the intermediate states $\ket{e'}$ and $\ket{d}$. \textbf{(c)} Most relevant off-resonant Rydberg states near $\ket{r_0}$ and $\ket{r_1}$. The thickness of the arrows indicates the transition weight of the corresponding transition.}
    \label{fig:atomic_transitions}
\end{figure}

\begin{table*}
\centering
\begin{minipage}{0.5\textwidth}
\centering
\renewcommand{\arraystretch}{1.2}
\begin{tabular}{c c c c c c}
\hline
$n$ & $R\,(\mathrm{\mu m})$ & $J/2\pi\,(\mathrm{MHz})$ & $V_{00}/J$ & $V_{01}/J$ & $V_{11}/J$\\
\hline
\multirow{3}{*}{$50$} & $10$ & $2.05$ & $7.5\times10^{-3}$ & $1.5\times10^{-3}$ & $6.0\times10^{-4}$\\
& $20$ & $0.256$ & $9.3\times10^{-4}$ & $1.9\times10^{-4}$ & $7.5\times10^{-5}$\\
& $30$ & $0.0759$ & $2.8\times10^{-4}$ & $5.7\times10^{-5}$ & $2.2\times10^{-5}$\\
\hline
\multirow{3}{*}{$75$} & $10$ & $11.2$ & $1.7\times10^{-1}$ & $4.0\times10^{-2}$ & $1.2\times10^{-2}$\\
& $20$ & $1.40$ & $2.2\times10^{-2}$ & $5.0\times10^{-3}$ & $1.5\times10^{-3}$\\
& $30$ & $0.414$ & $6.4\times10^{-3}$ & $1.5\times10^{-3}$ & $4.6\times10^{-4}$\\
\hline
\multirow{3}{*}{$100$} & $10$ & $36.7$ & $1.5$ & $4.0\times10^{-1}$ & $1.0\times10^{-1}$\\
& $20$ & $4.58$ & $1.9\times10^{-1}$ & $5.0\times10^{-2}$ & $1.3\times10^{-2}$\\
& $30$ & $1.36$ & $5.7\times10^{-2}$ & $1.5\times10^{-2}$ & $3.7\times10^{-3}$\\
\hline
\end{tabular}
\caption{Resonant dipole-dipole and vdW interaction
strengths, $J$ and $V_{ij}$, between pairs of Rydberg states $\ket{r_0}=\ket{nS_{1/2}, m_j=1/2}$ and $\ket{r_1}=\ket{nP_{3/2}, m_j=3/2}$ with $n=50, 75, 100$ for ${}^{87}$Rb at a distance $R$ \cite{S_Sibalic_2017}.}
\label{tab:atomic_data1}
\end{minipage}
\hfill
\begin{minipage}{0.45\textwidth}
\centering
\renewcommand{\arraystretch}{1.2}
\begin{tabular}{c|ccc}
 & $n=50$ & $n=75$ & $n=100$ \\
\hline
$C_3/2\pi\,(\mathrm{GHz\,\mu m^3})$ & $2.05$ & $11.2$ & $36.7$ \\
$C_6^{00}/2\pi\,(\mathrm{GHz\,\mu m^6})$ & $15$ & $1.9\times10^{3}$ & $5.6\times10^{4}$ \\
$C_6^{01}/2\pi\,(\mathrm{GHz\,\mu m^6})$ & $3.1$ & $4.5\times10^{2}$ & $1.5\times10^{4}$ \\
$C_6^{11}/2\pi\,(\mathrm{GHz\,\mu m^6})$ & $1.2$ & $1.4\times10^{2}$ & $3.7\times10^{3}$ \\
$\Gamma_0/2\pi\,(\mathrm{kHz})$ & $7.77$ & $2.15$ & $0.88$ \\
$\Gamma_1/2\pi\,(\mathrm{kHz})$ & $3.83$ & $1.08$ & $0.44$ \\
\hline
\end{tabular}
\vspace{32pt}
\caption{Dispersion coefficients of the resonant dipole-dipole and vdW interaction, $C_3$ and $C_6^{ij}$, and decay rates $\Gamma_j$ of Rydberg states $\ket{r_0}=\ket{nS_{1/2}, m_j=1/2}$ and $\ket{r_1}=\ket{nP_{3/2}, m_j=3/2}$ with $n=50, 75, 100$ for ${}^{87}$Rb \cite{S_Sibalic_2017}.}
\label{tab:atomic_data2}
\end{minipage}
\end{table*}

\subsection{Multi-photon transitions}

The protocol presented in the main text relies on the ability to realize the two atomic transitions $\ket{j}\leftrightarrow\ket{r_j}$ with constant Rabi frequency amplitudes $|\Omega_j|$ and optimally shaped, time-dependent Rabi frequency phases $\phi_j(t)$, for $j=0,1$. Two possible physical implementations of these transitions, together with the relevant atomic states, light frequencies, and polarizations, are shown in Figs.~\hyperref[fig:atomic_transitions]{\ref{fig:atomic_transitions}(a)-(b)} and discussed below.\\
Given the choice of states in the previous section, the direct transition $\ket{0}\leftrightarrow\ket{r_0}$ is forbidden by selection rules, since $\Delta l = 0$, where $l$ denotes the orbital angular momentum quantum number. The Rydberg state $\ket{r_0}$ can instead be accessed via a two-photon transition through an intermediate state $\ket{e}=\ket{6 P_{3/2}, F=2, m_F=1}$, or alternatively $\ket{e'}=\ket{5 P_{3/2}, F=2, m_F=1}$ (see Fig.~\hyperref[fig:atomic_transitions]{\ref{fig:atomic_transitions}(a)}). Scattering from the short-lived intermediate state $\ket{e}$ can be suppressed by employing a large intermediate detuning $\Delta_e$ and by maximizing the population of the so-called ``dark" state ($\ket{D}\propto-\ket{0}+\ket{r_0}$), which does not contain $\ket{e}$. The latter condition is achieved through an appropriate choice of the relative signs of the intermediate and two-photon detunings \cite{S_Boller_1991, S_Moller_2008}, and has already been experimentally demonstrated \cite{S_Evered_2023}.\\
We consider a three-level cascade system $\{\ket{0},\ket{e},\ket{r_0}\}$ and two lasers addressing the transitions $\ket{0}\leftrightarrow\ket{e}$ and $\ket{e}\leftrightarrow\ket{r_0}$, with Rabi frequencies $\Omega_0^{0e}$ and $\Omega_0^{e r_0}$ and detunings $\Delta_0^{0e}$ and $\Delta_0^{e r_0}$, respectively. The Hamiltonian is
\begin{equation}
    \label{H_2p}
    H_{2p}= 
    \begin{pmatrix}
        0 & \Omega_0^{0e}/2 & 0\\
        {\Omega_0^{0e}}^*/2 & \Delta_0^{0e} & \Omega_0^{e r_0}/2 \\
        0 & {\Omega_0^{e r_0}}^*/2 & \Delta_0^{0e}+\Delta_0^{e r_0}
    \end{pmatrix}.
\end{equation}
In the limit of large intermediate detuning, $\Delta_e\equiv\Delta_0^{0e} \gg \Omega_0^{0e}, \Omega_0^{e r_0}$, the population of the intermediate state $\ket{e}$ is strongly suppressed, and the two-photon Hamiltonian $H_{2p}$ in Eq.~\eqref{H_2p} can be reduced to the following effective two-level Hamiltonian in the subspace ${\ket{0},\ket{r_0}}$
\begin{equation}
    H_{2p}^{\rm eff}=
    \begin{pmatrix}
        \delta_0 & \Omega_0^{\rm eff}/2\\
        {\Omega_0^{\rm eff}}^*/2 & \delta_{r_0}+\Delta_e+\Delta_0^{e r_0}
    \end{pmatrix},
\end{equation}
where the effective two-photon Rabi frequency is $\Omega_0^{\mathrm{eff}} = -\Omega_0^{0e}\Omega_0^{e r_0}/(2\Delta_e)$, the Stark shifts of the states $\ket{0}$ and $\ket{r_0}$ are $\delta_0 = -|\Omega_0^{0e}|^2/(4\Delta_e)$ and $\delta_{r_0} = -|\Omega_0^{e r_0}|^2/(4\Delta_e)$, respectively, and the resonance condition is $\Delta_0^{e r_0} = -\Delta_e + \delta_0 - \delta_{r_0}$. Thus, the phases of the two lasers must satisfy $\phi_0^{0e}(t) + \phi_0^{e r_0}(t) = \phi_0(t)$, where $\phi_0 (t)$ denotes the optimal phase profile.\\
As shown in \cite{S_Evered_2023}, once the resonance condition is met, scattering due to the intermediate state can be further reduced by exploiting the dark state. The Rydberg population is predominantly realized via the dark state when the intermediate-state detuning $\Delta_e$ and the two-photon detuning $\delta$ — which is determined by the time dependence of $\Omega_0^{\rm eff}$, i.e., $\delta\propto-\dot{\phi_0}$ — have opposite signs, i.e., $\delta\Delta_e<0$. For a time-dependent detuning $\delta(t)$, the relevant sign is determined by its value at the beginning of the pulse, $\delta(0)\propto -\dot{\phi}_0(0)$. This situation occurs, for example, for the pulses presented in the main text, where the optimal phase profiles $\phi_j(t)$ exhibit a time-dependent derivative, thus a time.\\
Note that the Stark shift $\delta_{0}$ leads to the accumulation of a dynamical phase, resulting in a relative phase between the states $\ket{0}$ and $\ket{1}$. This effect can be easily compensated by modifying the final single-qubit rotation $R(\theta, \varphi, \lambda)$, without altering the optimal pulse itself.\\

Since the states $\ket{1}$ and $\ket{r_1}$ have different parity, the transition $\ket{1}\leftrightarrow\ket{r_1}$ can be implemented using a single laser field, in particular a UV laser. While working with high-power vacuum ultraviolet radiation presents significant experimental challenges, it remains a feasible approach \cite{S_Hankin_2014, S_Jau_2016, S_Saffman_2016, S_Li_2019, S_Ma_2023}. If the single-photon UV scheme is impractical, a three-photon transition using two intermediate states, such as $\ket{e'}$ and $\ket{d}=\ket{5 D_{5/2}, F=3, m_F=2}$, can still be implemented, as shown in Fig.~\hyperref[fig:atomic_transitions]{\ref{fig:atomic_transitions}(b)}. The effective three-photon Rabi frequency is then $\Omega_1^{\rm eff} \approx \Omega_1^{0e'}\Omega_1^{e'd}\Omega_1^{dr_1}/(2{\Delta_1^{1e'}})^2$. In the following, we consider a Rydberg excitation scheme that employs a UV laser for the transition $\ket{1}\leftrightarrow\ket{r_1}$, as in Fig.~\hyperref[fig:atomic_transitions]{\ref{fig:atomic_transitions}(a)}.

\subsection{Other off-resonant Rydberg states}

In realistic Rydberg excitation pulses the laser field also induces off-resonant couplings to Rydberg states other than the intended target state $\ket{r}$. Given an initial state $\ket{a}$ — that is, $\ket{a}=\ket{e\,(1)}$ for the target Rydberg state $\ket{r}=\ket{r_0\,(r_1)}$ — the most relevant off-resonant Rydberg states $\ket{r'}$ coupled to $\ket{a}$ by a light of polarization $q$ can be identified by computing the detuning $\Delta=|E_{r'}-E_a-\omega|=|E_{r'}-E_{r}|$, where $E_r$ denotes the energy of the state $\ket{r}$ and $\omega=E_{r}-E_a$ the frequency of the laser, together with the corresponding transition dipole moment $d=|\bra{r'}d^q\ket{a}|$ for all states in the vicinity of the target Rydberg state $\ket{r}$. These states can then be ranked according to a transition weight $W$ inspired by second-order perturbation theory and normalized using $d_{\rm targ}=|\bra{r}d^q\ket{a}|$, the transition dipole moment of the desired transition.
\begin{equation}
    \label{TW}
    W(r, r') = 
    \begin{cases}
        \dfrac{d^2}{d_{\rm targ}\Delta} & \text{if}\, \Delta\neq0\\
        1 & \text{otherwise}        
    \end{cases}.
\end{equation}
Figure~\hyperref[fig:atomic_transitions]{\ref{fig:atomic_transitions}(c)} and Tab.~\ref{tab:off_resonant} show the dominant off-resonant transitions that can affect the two Rydberg excitations $\ket{0}\leftrightarrow\ket{r_0}$ and $\ket{1}\leftrightarrow\ket{r_1}$ for a principal quantum number $n = 100$. In particular, the dominant contribution for the former arises from the state $\ket{r_0'} = \ket{98D_{5/2}, m_J = -3/2}$ (Fig.~\hyperref[fig:atomic_transitions]{\ref{fig:atomic_transitions}(c.1)}), while for the latter it comes from $\ket{r_1'} = \ket{100D_{3/2}, m_J = 1/2}$ (Fig.~\hyperref[fig:atomic_transitions]{\ref{fig:atomic_transitions}(c.1)}). At zero magnetic field, $B = 0$, the states $\ket{r_1}$ and $\ket{r_1'}$ are degenerate; therefore, a sufficiently strong magnetic field $B$ is required to spectrally resolve the transition, even in the presence of a strong driving field, such that $\Omega_1\ll \mu_B g_J B$, where $\mu_B$ is the Bohr magneton and $g_J\approx1.3$ is the Landé factor for the $P_{3/2}$ state.

\begin{table*}[t]
\centering
\renewcommand{\arraystretch}{1.2}
\begin{tabular}{c|c|c|c|c }
\hline
target Rydberg state $\ket{r}$ & off-resonant Rydberg state $\ket{r'}$ & $\Delta$ ($\mathrm{GHz}$) & $d$ ($ea_0$) & $W$ ($ea_0/\mathrm{GHz}$)\\
\hline
\multirow{5}{*}{$\ket{r_0}=\ket{100S_{1/2},m_j=1/2}$} 
& $\ket{100S_{1/2},m_j=1/2}$ & $0$ & $0.0037$ & $1$\\
& $\ket{98D_{5/2},m_j=-3/2}$ & $1.8$ & $0.0047$ & $0.0035$\\
& $\ket{99D_{5/2},m_j=-3/2}$ & $5.4$ & $0.0047$ & $0.0011$\\
& $\ket{97D_{5/2},m_j=-3/2}$ & $9.2$ & $0.0048$ & $0.00069$\\
& $\ket{98D_{5/2},m_j=1/2}$ & $1.6$ & $0.0019$ & $0.00064$\\
\hline
\multirow{3}{*}{$\ket{r_1}=\ket{100P_{3/2},m_j=3/2}$} 
& $\ket{100P_{3/2},m_j=3/2}$ & $0$ & $0.00046$ & $1$\\
& $\ket{100P_{3/2},m_j=1/2}$ & $0.093$ & $0.00027$ & $0.0012$\\
& $\ket{100P_{1/2},m_j=1/2}$ & $0.21$ & $0.00029$ & $0.00087$\\
\hline
\end{tabular}
\caption{Off-resonant Rydberg states with their detuning $\Delta$, dipole moment $d$ and transition weight $W$ (see Eq.~\eqref{TW}), with a magnetic field $B=50\,\mathrm{G}.$
}
\label{tab:off_resonant}
\end{table*}

\section{Dipole-dipole interaction}

The dipole–dipole interaction $V_{\rm dd}(\vec{R})$ between two dipoles $\vec{d}_\alpha$ ($\alpha = A, B$) separated by a distance vector $\vec{R}$ reads \cite{S_Chomaz_2023, S_Emperauger_2025}
\begin{equation}
    \label{V_dd}
    V_{\rm dd}(\vec{R}) = \frac{\vec{d}_A \cdot \vec{d}_B - 3(\vec{d}_A \cdot \hat{R})(\vec{d}_B \cdot \hat{R})}{R^3},
\end{equation}
where $R = |\vec{R}|$ is the magnitude of the interdipole separation, and $\hat{R} = \vec{R}/R$ is the corresponding unit vector.\
The expression in Eq.~\eqref{V_dd} can also be rewritten as \cite{S_Wall_2015, S_Chomaz_2023}
\begin{align}
    \nonumber
    V_{\rm dd}(\vec{R}) = & \frac{1}{R^3}\Bigg[ \frac{1-3\cos^2\theta}{2} (d_{A}^{+1}d_{B}^{-1}+d_{A}^{-1}d_{B}^{+1}+2d_{A}^{0}d_{B}^{0})\\
    \nonumber
    & + \frac{3}{\sqrt{2}} \sin\theta\cos\theta\left((d_{A}^{+1}d_{B}^{0}+d_{A}^{0}d_{B}^{+1})e^{-i\varphi} - (d_{A}^{-1}d_{B}^{0}+d_{A}^{0}d_{B}^{-1})e^{i\varphi}\right)\\
    \label{V_dd2}
    & - \frac{3}{2}\sin^2\theta \left( d_{A}^{-1}d_{B}^{-1}e^{2i\varphi} + d_{A}^{+1}d_{B}^{+1}e^{-2i\varphi} \right)
    \Bigg],
\end{align}
where the separation vector $\vec{R}$ has been expressed in spherical coordinates as $\vec{R} = R (\sin\theta\cos\varphi, \sin\theta\sin\varphi, \cos\theta)$, with $\theta$ the polar angle between $\vec{R}$ and the $z$ axis, and $\varphi$ the azimuthal angle. The dipole moment operators $\vec{d}_\alpha = (d^x_\alpha, d^y_\alpha, d^z_\alpha)$ are also expressed in terms of their spherical basis components $\vec{d}_\alpha = (d^{+1}_\alpha, d^{-1}_\alpha, d^0_\alpha)$ using the following transformation rules
\begin{equation}
    \begin{cases}
        d^x_\alpha = \dfrac{d_{\alpha}^{-1}-d_{\alpha}^{+1}}{\sqrt{2}}\\
        d^y_\alpha = i\dfrac{d_{\alpha}^{-1}+d_{\alpha}^{+1}}{\sqrt{2}}\\
        d^z_\alpha = d^0_\alpha
    \end{cases}
\end{equation}
One can verify that, given the choice of states described in the previous section, the only nonvanishing transition dipole moments are
\begin{equation}
    \bra{r_0} d^{-1}_\alpha \ket{r_1} = \bra{r_1} d^{+1}_\alpha \ket{r_0} \equiv d,
\end{equation}
since the dipole matrix elements in the fine-structure basis $\ket{n, l, j, m_j}$ — with $n$ the principal quantum number, $l$ the electronic orbital angular momentum quantum number, $j$ the total electronic angular momentum quantum number, and $m_j$ its projection onto the quantization axis $z$ — are given by \cite{S_Sibalic_2017}
\begin{equation}
    \bra{n,l,j,m_j}d^q_\alpha\ket{n',l',j',m_j'} \propto
    \begin{pmatrix}
        l & 1 & l'\\
        0 & 0 & 0
    \end{pmatrix}
    \begin{pmatrix}
        j & 1 & j'\\
        -m_j & -q & m_j'
    \end{pmatrix}.
\end{equation}
The selection rules imposed by the Wigner 3-$j$ symbols require $l + l' + 1$ to be even, implying that the two states must have opposite parity, and that the polarization $q$ must satisfy $-m_j - q + m_j' = 0$.\\
Moreover, the only relevant terms in Eq.~\eqref{V_dd2} are those responsible for the resonant exchange interaction, i.e., $d_{A}^{\pm1} d_{B}^{\mp1}$. Hence, the transition matrix elements of the resonant exchange dipole–dipole interaction are
\begin{equation}
    \label{J(R)}
    J(\vec{R}) = \bra{r_0r_1} V_{\rm dd}(\vec{R}) \ket{r_1r_0} = \bra{r_1r_0} V_{\rm dd}(\vec{R}) \ket{r_0r_1} = \frac{d^2}{R^3} \frac{1-3\cos^2\theta}{2}.
\end{equation}
In this work, we consider a quantization axis $z$ aligned with the interatomic direction $\vec{R}$, thus $\theta = 0$. In this way, we can write
\begin{equation}
    J(R) = \frac{C_3}{R^3}
\end{equation}
with $C_3 = -d^2=-|\bra{r_0}d_{-1}^{\alpha}\ket{r_1}|^2$ the dispersion coefficient of the dipole-dipole interaction.

\section{van der Waals interactions}

At large interatomic separations, the van der Waals (vdW) interaction represents a small perturbative correction to the resonant dipole–dipole coupling $J$, and decays as $V_{\rm vdW} = C_6 / R^6$, where $C_6$ is the dispersion coefficient associated with the vdW forces and depends on the specific pair state $\ket{r_i r_j}$ \cite{S_Saffman_2010}. Within second-order perturbation theory, $C_6$ can be expressed as \cite{S_Sibalic_2017}
\begin{equation}
    C^{ij}_6 = \sum_{r', r''} \frac{\left|\bra{r'r''} V_{\rm dd} \ket{r_ir_j} \right|^2}{\Delta_{r'r''}^{ij}},
\end{equation}
where $V_{\rm dd}$ is the dipole–dipole interaction operator defined in Eq.~\eqref{V_dd}, and the sum runs over all intermediate Rydberg states $r'$ and $r''$, given a sufficiently small detuning $\Delta_{r'r''}^{ij}=E_{r'r''}-E_{r_ir_j}$, with $E_{r_ir_j}$ the energy of the pair state $\ket{r_i r_j}$. To avoid divergences in the perturbative expression, when evaluating $C^{01}_6$ the symmetric state $\ket{r_1 r_0}$ must be excluded from the sum, as it is resonant with $\ket{r_0 r_1}$ and its contribution is already explicitly accounted for in the resonant exchange interaction. The same consideration applies to the computation of $C^{10}_6$. Since the vdW dispersion coefficients scale as $C_6 \sim n^{11}$, the ratio between the vdW interaction and the resonant dipole–dipole coupling scales as $V_{\rm vdW}/J \sim n^7 / R^3$. While in the long-distance regime — the primary focus of this work — the vdW contribution becomes less significant, the large principal quantum numbers $n\sim100$ required to achieve long Rydberg lifetimes make these interactions non-negligible (see Tab.~\ref{tab:atomic_data1}).
 
\section{ \label{sec:protocol_piJpi} \texorpdfstring{$\pi J \pi$}{pi J pi} protocol  }

A simple scheme to implement the iSWAP gate with Rydberg atoms — referred to as the $\pi J \pi$ protocol and inspired by molecular platforms \cite{S_Ni_2018, S_Holland_2023, S_Bao_2023, S_Picard_2025, S_Ruttley_2025} — consists of three steps:
\begin{enumerate}
    \item \textit{Toggling $\pi$ pulses}: The fields $\Omega_j$ drive the transitions $\ket{j} \rightarrow \ket{r_j}$ ($j=0,1$), transferring the population from the ground-states to the Rydberg states and thereby activating the dipolar interaction. The duration is $T_\pi = \pi / \Omega_{\text{max}}$;
    \item \textit{Free dipole–dipole interaction}: By waiting for a time $T_J = \pi/(2J)$, the spin-exchange Hamiltonian $H_J$ generates a maximally entangled state;
    \item \textit{De-toggling $\pi$ pulses}: The fields $\Omega_j$ drive the transitions $\ket{r_j} \rightarrow \ket{j}$ ($j=0,1$), returning the population from the Rydberg states to the ground-states. The duration is $T_\pi$.
\end{enumerate}
The total duration $T_{\pi J \pi}$ of the $\pi J \pi$ sequence  is
\begin{equation}
    T_{\pi J \pi}\Omega_{\rm max} = (2 T_\pi + T_J)\Omega_{\rm max} = \pi\left(\frac{4+\Omega_{\text{max}}/J}{2}\right).
\end{equation}
For a reasonable choice of Rabi frequency, e.g., $\Omega_{\text{max}}/2\pi=1\,\mathrm{MHz}$, and choosing the principal quantum number $n=100$ and the interatomic distance $R=11\,\mathrm{\mu m}$ — one has $\Omega_{\text{max}}/J = 0.1$, resulting in $T_{\pi J \pi} = 6.4 / \Omega_{\text{max}}$, which is approximately 15\% faster than the time-optimal gate based on the blockade mechanism \cite{S_Jandura_2022}. This is promising, but there is a caveat. When $\Omega_{\text{max}} \lesssim J$, the dipolar interaction $J$ significantly affects the dynamics even during the toggling $\pi$ pulses. Indeed, simulating the $\pi J \pi$ protocol with $\Omega_{\text{max}}/J = 0.1$ reveals that the fidelity $F$ is very low, as shown in Fig.~\hyperref[fig:miscellanea]{\ref{fig:miscellanea}(a)}. Achieving reasonable fidelity requires a much larger ratio $\Omega_{\text{max}}/J$ to neglect the effect of dipolar exchange during the toggling pulses — as is the case for molecules. However, increasing this ratio also prolongs the gate duration. For example, for $\Omega_{\text{max}}/J = 50$, the infidelity is $1-F \approx 10^{-3}$, but the gate duration becomes $T_{\pi J \pi}\Omega_{\text{max}} = 85$, which is excessively long, especially considering the finite lifetimes of Rydberg states, as illustrated in Fig.~\hyperref[fig:miscellanea]{\ref{fig:miscellanea}(b)}, where the effects of spontaneous emission are included in the simulations.\\

\section{Gate with constant pulses}

\begin{figure*}[t]
    \centering
    \includegraphics[width=0.95\linewidth]{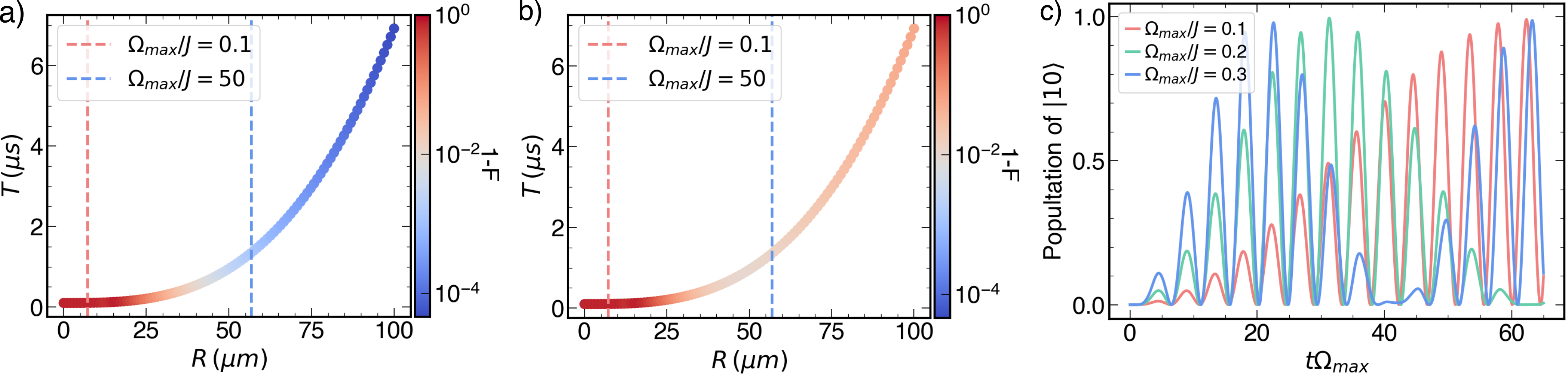}
    \caption{\textbf{(a)} Simulation of the fidelity $\pi J \pi$ protocol with $\Omega_{\text{max}}/2\pi=10\,\mathrm{MHz}$, and $n=100$, in absence of decay from the Rydberg states. Each point represents a different ratio $\Omega_{\text{max}}/J$, thus a different gate duration $T$ and range $R$. \textbf{(b)} Same simulation of Fig.~\hyperref[fig:miscellanea]{\ref{fig:miscellanea}(a)}, including also the Rydberg decay, $\Gamma_0/2\pi=0.88\,\mathrm{kHz}$ and $\Gamma_1/2\pi=0.44\,\mathrm{kHz}$. \textbf{(c)} Time evolution of the population of the state $\ket{10}$ when starting from $\ket{01}$ and applying constant global pulses $\Omega_j(t)=\Omega_{\text{max}}$ ($j=0,1$) with three different ratios $\Omega_{\text{max}}/J$.}
    \label{fig:miscellanea}
\end{figure*}

In the main text, it was shown that the time-optimal $U_{\rm iSWAP}$ gate can be implemented using a pulse with an optimized phase. In this section, we show that the same unitary $U_{\rm iSWAP}$ can also be realized by applying a single global constant pulse of duration $T$, without any phase modulation — although the resulting gate is considerably slower than the time-optimal one. The duration $T$ is chosen such that, by the end of the pulse, the entire iSWAP operation is completed: the population is transferred to the Rydberg states, the dipolar exchange occurs, and the population returns to the ground states — all within a single step.\\
In Fig.~\hyperref[fig:miscellanea]{\ref{fig:miscellanea}(c)}, the system is initialized in the state $\ket{01}$, after which constant pulses $\Omega_j(t)=\Omega_{\text{max}}$ ($j=0,1$) are applied. The figure shows the time evolution of the population of $\ket{10}$, allowing us to identify when a complete transfer $\ket{01}\rightarrow\ket{10}$ occurs. This is evaluated for different values of the ratio $\Omega_{\text{max}}/J$. A full population transfer is observed, for instance, at $\Omega_{\text{max}}/J=0.1$ with the “magic” gate time $T\Omega_{\text{max}}=62.3$, and at $\Omega_{\text{max}}/J=0.2$ with $T\Omega_{\text{max}}=31.2$. More generally, the “magic" gate duration $T$ yielding complete transfer scales approximately as the inverse of $\Omega_{\text{max}}/J$, at least for $\Omega_{\text{max}}/J<1$. For larger ratios, however, the maximum of the transferred population decreases: for example, in Fig.~\hyperref[fig:miscellanea]{\ref{fig:miscellanea}(c)}, the first major peak at $\Omega_{\text{max}}/J=0.3$ occurs earlier than for $\Omega_{\text{max}}/J=0.1$, but reaches a smaller maximum and does not achieve unity. To reach lower infidelities and time-optimal performance, it is necessary to optimize the pulse phases.

\section{GRAPE}

The GRAPE (Gradient Ascent Pulse Engineering) algorithm is a standard optimal control technique that finds controls $\vec{u}(t)$ minimizing a cost functional $\mathcal{C}[\vec{u}(t)]$ for a Hamiltonian $H(\vec{u}(t))$ \cite{S_Khaneja_2005, S_Jandura_2022}. Here, the BFGS method is used \cite{S_SciPy_2020}. A piecewise-constant ansatz divides the pulse duration $T$ into $N$ steps of length $\Delta t = T/N$ and assuming that within each time-step each control has a constant value, i.e., $\vec{u}(t) = \vec{u}^k$ for $t \in [k\Delta t, (k+1)\Delta t]$ and $k=0,1,...,N-1$.\\
In the system of two dipole–dipole–interacting atoms driven by two Rydberg excitation fields, the available controls are the Rabi amplitudes $|\Omega_j(t)|$ and phases $\phi_j(t)$ ($j=0,1$). We set $|\Omega_0(t)| = |\Omega_1(t)| = \Omega_{\text{max}}$ and $\phi_0(t) = \phi_1(t) = \phi(t)$. This choice simplifies the optimization without affecting the time-optimality of the resulting pulses. Indeed, even when this assumption is relaxed, we find that GRAPE naturally favors equal phases and, at least for relatively small ratios $\Omega_{\text{max}}/J$, drives the Rabi frequencies to their maximum value and keeps them constant throughout the pulse. For large ratios $\Omega_{\text{max}}/J \gtrsim 10$, it is no longer necessary to keep the driving fields on throughout the entire gate, and GRAPE tends to converge to solutions resembling the $\pi J \pi$ protocol, with two laser pulses at the beginning and end of the gate, separated by a period of free dipole–dipole interaction. Three additional parameters, $\theta$, $\varphi$, and $\lambda$, account for a final global single-qubit rotation $R(\theta,\varphi,\lambda) = \exp(i(\varphi+\lambda)/2) R_z(\varphi) R_y(\theta) R_z(\lambda)$. The full set of optimization parameters thus reads $\vec{u} = \{\phi^k, \theta, \varphi, \lambda\}^{k=0,\dots,N-1}$.

\begin{figure*}[t]
    \centering
    \includegraphics[width=0.9\linewidth]{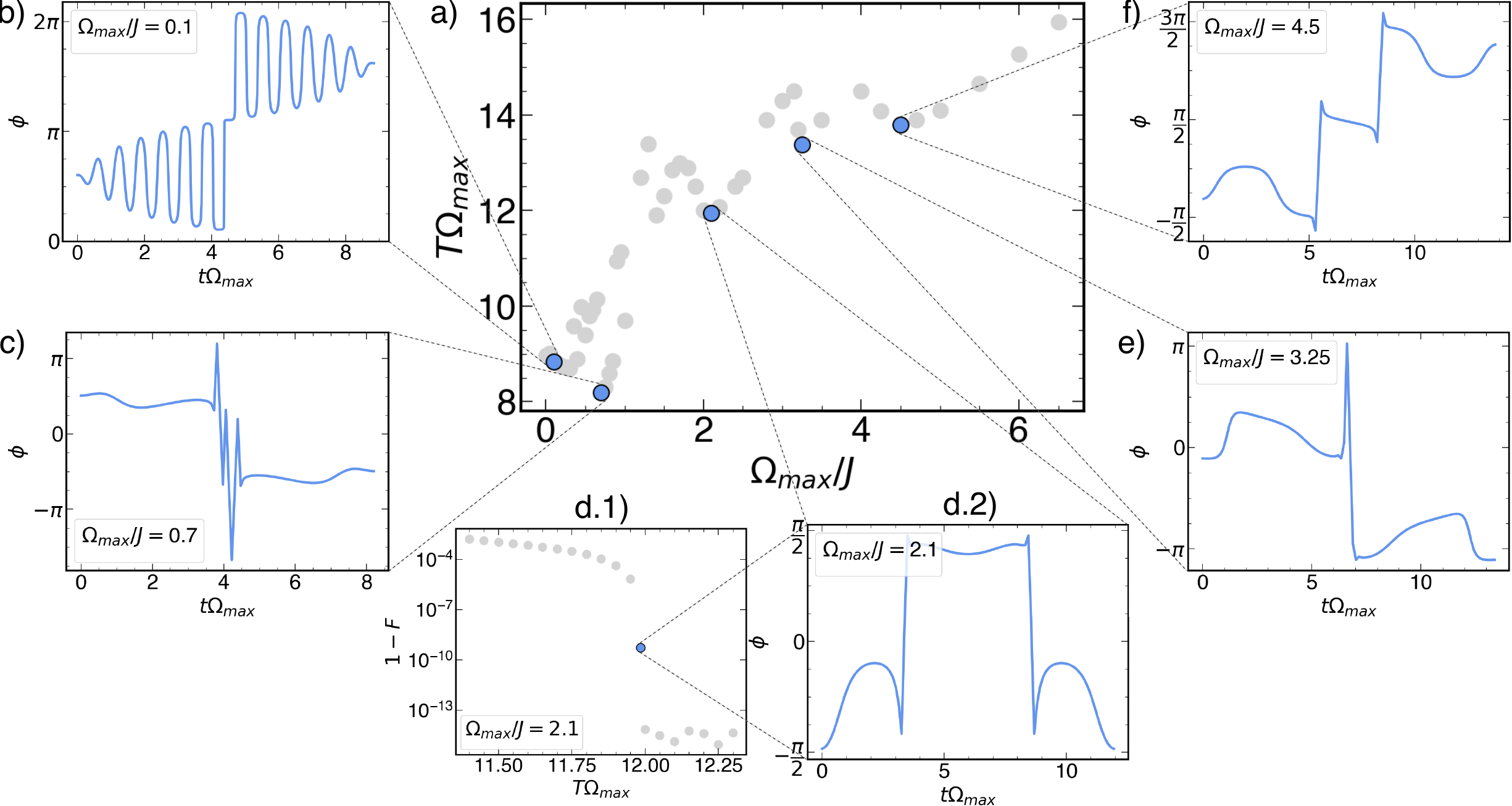}
    \caption{Time-optimal pulses. \textbf{(a)} The time-optimal gate duration $T\Omega_{\text{max}}$ for different ratios $\Omega_{\text{max}}/J$. \textbf{(b)} Phase $\phi(t)$ of the time-optimal pulse for $\Omega_{\text{max}}/J=0.1$, with duration $T\Omega_{\text{max}}=8.84$. \textbf{(c)} Phase $\phi(t)$ of the time-optimal pulse for $\Omega_{\text{max}}/J=0.7$, with duration $T\Omega_{\text{max}}=8.2$. \textbf{(d.1)} Infidelity $1-F$ as a function of the gate time duration $T\Omega_{\text{max}}$ for the ratio $\Omega_{\text{max}}/J=2.1$. In order to achieve a numerical zero infidelity pulses of duration $T\Omega_{\text{max}}\ge 11.95$ are required. \textbf{(d.2)} Phase $\phi(t)$ of the time-optimal pulse for $\Omega_{\text{max}}/J=2.1$, with duration $T\Omega_{\text{max}}=11.95$. \textbf{(e)} Phase $\phi(t)$ of the time-optimal pulse for $\Omega_{\text{max}}/J=3.25$, with duration $T\Omega_{\text{max}}=13.38$. \textbf{(f)} Phase $\phi(t)$ of the time-optimal pulse for $\Omega_{\text{max}}/J=4.5$, with duration $T\Omega_{\text{max}}=13.8$. }
    \label{fig:S_pulses}
\end{figure*}

\section{\label{sec:results} Time-optimal pulses}

In Fig.~\ref{fig:S_pulses}, the time-optimal phase profiles $\phi(t)$ obtained with GRAPE for different values of the ratio $\Omega_{\text{max}}/J$ are shown. In particular, Fig.~\hyperref[fig:S_pulses]{\ref{fig:S_pulses}(a)} displays the time-optimal gate duration $T\Omega_{\text{max}}$ as a function of $\Omega_{\text{max}}/J$. The quantity $T\Omega_{\text{max}}$ represents the minimum pulse duration for which GRAPE identifies a pulse yielding a (numerically) vanishing cost function, i.e., $1-F \approx 0$ [see Fig.~\hyperref[fig:S_pulses]{\ref{fig:S_pulses}(d.1)}].\\
When $\Omega_{\text{max}}/J < 1$, the time-optimal pulses display similar features [see Fig.~\hyperref[fig:S_pulses]{\ref{fig:S_pulses}(b)}]. In particular, the phase $\phi(t)$ exhibits damped oscillations at frequency $J$. The Fourier spectra of the pulses show a dominant central peak at the laser resonance frequency, accompanied by additional peaks at frequencies $\pm nJ$ (with odd $n$). These sidebands, detuned by $\pm nJ$ from resonance, effectively compensate for the energy shifts induced by the dipole–dipole interaction. Moreover, a phase jump of approximately $\pi$ typically occurs near the middle of the pulse.\\
When the ratio $\Omega_{\text{max}}/J$ approaches or exceeds unity, the behavior of the phase $\phi(t)$ becomes less regular, although the characteristic phase jump near the middle of the pulse generally persists [see Figs.~\hyperref[fig:S_pulses]{\ref{fig:S_pulses}(c)–(f)}].\\
The shortest pulse is obtained for $\Omega_{\text{max}}/J = 0.7$, with $T\Omega_{\text{max}} = 8.2$ [see Fig.~\hyperref[fig:S_pulses]{\ref{fig:S_pulses}(c)}].
Relatively short durations are also found for $\Omega_{\text{max}}/J = 2.1$ ($T\Omega_{\text{max}} = 11.95$), $\Omega_{\text{max}}/J = 3.25$ ($T\Omega_{\text{max}} = 13.38$), and $\Omega_{\text{max}}/J = 4.5$ ($T\Omega_{\text{max}} = 13.8$) [see Figs.~\hyperref[fig:S_pulses]{\ref{fig:S_pulses}(d.2)–(f)}].

\section{Simulation of motion}

To obtain realistic estimates of the fidelity of the optimal pulses, it is necessary to include the motional degrees of freedom of the atoms. These can affect the dipole–dipole interaction by altering the interatomic distance, or induce decoherence by generating entanglement between internal and motional states. In this section, we discuss the kinetic Hamiltonian, the effects of atomic displacements on the interaction strength, and the impact of photon recoil. For each contribution, we estimate its effect on the pulse infidelity through numerical simulations and analytical calculations of the corresponding coupling strengths.

\subsection{Kinetic energy}

The three-dimensional potential of the optical tweezers in which the atoms are trapped — prior to the Rydberg excitation pulses — can be modeled as three harmonic oscillators with angular frequencies $\omega_\ell$, where $\ell = x$ (axial direction) and $\ell = y,z$ (radial directions) [see Fig.~\hyperref[fig:S_fidelity]{\ref{fig:S_fidelity}(d)}]. Typically, $5\omega_x \approx \omega_y \approx \omega_z$. The kinetic Hamiltonian of the atoms, $H_K$, is given by
\begin{equation}
    \label{H_kinetic}
    H_K = - \sum_{\alpha=A,B} \frac{\vec{p}^2_\alpha}{2m},
\end{equation}
where $\vec{p}_\alpha=(p^x_\alpha, p^y_\alpha, p^z_\alpha)$, and $p^\ell_\alpha=\sqrt{m\omega_\ell/2}(a_{\ell,\alpha}^\dag-a_{\ell,\alpha})$ is the momentum operator of atom $\alpha$ along direction $\ell = x,y,z$. Here, $a_{\ell,\alpha}$ and $a_{\ell,\alpha}^\dag$ are the corresponding ladder operators. We omit the trapping potential, as optical tweezers are typically turned off during the Rydberg excitation pulses.

\subsection{\label{app:distance_fluctuations}Distance fluctuations}

Equation~\eqref{J(R)} describes the behavior of the exchange dipole-dipole interaction strength $J$ as a function of the vector distance $\vec{R}$ between two atoms. Namely, it depends on the magnitude of the distance $R$ and the polar angle $\theta$ between the vector $\vec{R}$ and the quantization axis $z$. In the following, we analyze how the atomic motion affects the interatomic distance $R$, the angle $\theta$, and consequently the coupling strength $J$.

\subsubsection{Variations of the radius \texorpdfstring{$R$}{}}

The position operator of molecule $\alpha$ with respect to the center of its optical trap — i.e., the minimum of the three trapping potentials — is $\vec{r}_\alpha = \left(x_\alpha,y_\alpha,z_\alpha\right)$, where $r_{\ell,\alpha} = \sqrt{1/(2m\omega_\ell)}(a_{\ell,\alpha}+a_{\ell,\alpha}^\dag)$ and $\ell=x,y,z$.\\
Consider the reference frame in which $x$ is the axial axis of the tweezers, and the interatomic separation vector $\vec{R}$ lies along the $z$ axis: $\vec{R}=(0,0,R)$ [see Fig.~\hyperref[fig:S_fidelity]{\ref{fig:S_fidelity}(d)}].\\
In the presence of motion — and even at zero temperature, due to the finite width of the motional ground-state wave function — computing the distance between atoms requires accounting for the position fluctuations $\vec{r}_\alpha$. The perturbed intermolecular separation vector thus reads $\vec{R}' = \vec{R} + \vec{r}_A - \vec{r}_B$. For sufficiently strong trapping potentials $\omega_\ell$ and sufficiently low temperature $T_{\rm temp}$, the relative dimensionless deviations $\epsilon_{\ell} = (r_{\ell,A} - r_{\ell,B})/R$ are small and can be treated as perturbative parameters. Hence, the magnitude of the perturbed distance is
\begin{align}
    R'= R \sqrt{\epsilon_x^2+\epsilon_y^2+(1+\epsilon_z)^2} = R\sqrt{1+\epsilon},
\end{align}
where $\epsilon$ is the collective perturbative parameter that takes into account the motion in all directions, its explicit expression is
\begin{equation}
    \epsilon=2\epsilon_z+\epsilon_x^2+\epsilon_y^2+\epsilon_z^2.
\end{equation} 
The dipole-dipole perturbed interaction strength $J'$ depends on the inverse of the cubic of $R'$ (see Eq.~\eqref{J(R)}), then we can expand it up to second order in $\epsilon$
\begin{equation}
    \label{1/R^3}
    \frac{1}{R'^3} = \frac{1}{R^3} \left( 1-\frac{3}{2}\epsilon+\frac{15}{8}\epsilon^2+o(\epsilon^3)\right).
\end{equation}

\subsubsection{Variations of the angle \texorpdfstring{$\theta$}{}}

Given the reference frame in which $x$ is the axis of the tweezers and the intermolecular separation vector $\vec{R}$ lies along the $z$ axis [see Fig.~\hyperref[fig:S_fidelity]{\ref{fig:S_fidelity}(d)}], the angle $\theta$ in Eq.~\eqref{J(R)} is zero, $\theta = 0$. Once the perturbations $\vec{r}_j$ are introduced, the perturbed angle $\theta'$ is no longer zero. The angular deviation $\theta'$ can be computed by projecting the vector $\vec{R}'$ onto the $z$ axis (unit vector $\hat{z}$), thus
\begin{equation}
    \cos\theta' = \frac{\vec{R}'\cdot\hat{z}}{R'} = \frac{1+\epsilon_z}{\sqrt{1+\epsilon}}.
\end{equation}
The angular part of the perturbed coupling strength $J'$ in Eq.~\eqref{J(R)} becomes
\begin{align}
    \label{theta'}
    \frac{1-3\cos^2\theta'}{2} & = \frac{1}{2}\left(1- 3 \frac{(1+\epsilon_z)^2}{1+\epsilon}\right) =-1 + \frac{3}{2}(\epsilon_x^2 +\epsilon_y^2) + o(\epsilon^3).
\end{align}

\subsubsection{Variations of the coupling strength}

Combining the effects of motion on both the intermolecular distance $R$ and the angle $\theta$, Eqs.~\eqref{1/R^3} and \eqref{theta'}, one can estimate the perturbed coupling constant $J'$ as
\begin{align}
    J'= \frac{d^2}{R'^3}\frac{1-3\cos^2\theta'}{2} = J \left( 1 - 3 \epsilon_z + 3(2\epsilon_z^2 - \epsilon_x^2 - \epsilon_y^2)\right)+o(\epsilon^3),
\end{align}
where, in the last line, we used the fact that in this frame $J = -d^2 / R^3$. Therefore, the exchange interaction Hamiltonian $H'_J$, accounting for the effect of atomic motion, becomes
\begin{align}
    \label{noise_J}
    H_{J}' = &  J \Bigg[1 - 3\frac{z_A-z_B}{R} + 6 \frac{(z_A-z_B)^2}{R^2} - 3\frac{(x_A-x_B)^2}{R^2} - 3\frac{(y_A-y_B)^2}{R^2} + ... \Bigg] \left(\ket{r_0r_1}\bra{r_1r_0}+\text{h.c.}\right)\\
    \label{H'_J}
    = & H_J + H_J^{(1,z)} + H_J^{(2,z)} + H_J^{(2,x)} + H_J^{(2,y)} + ...
\end{align}
where $H_J^{(n,\ell)}$ refers to the motion-noise-term at the perturbative order $n$ relative to the direction $\ell=x,y,z$.\\

Similarly, for the van der Waals interactions $V_{\rm vdW} \sim R^{-6}$, one can demonstrate that the Hamiltonian in the presence of distance fluctuations at first-order reads
\begin{align}
    H_{\rm vdW}' & =  H_{\rm vdW} + H_{\rm vdW}^{(1,z)}\\
    & = \sum_{i,j} V_{ij} \left[1-6\frac{z_A-z_B}{R}\right] \ket{r_i r_j} \bra{r_i r_j} 
\end{align}
where $H_{\rm vdW}$ is the interaction Hamiltonian in absence of distance fluctuations and $H_{\rm vdW}^{(1,z)}$ refers to the motion-noise-term at the first perturbative order relative to the direction $\ell=z$. Higher-order terms are not required, as even the first-order term does not significantly affect the dynamics (see Tab.~\ref{tab:noisy_terms}).

\subsection{Photon recoil}

When an atom absorbs a photon, it also acquires the corresponding momentum $k=2\pi/\lambda$, where $\lambda$ is the wavelength of the incident light. This momentum kick perturbs the motional state of the atom and can be modeled by introducing the Lamb–Dicke parameter $\eta_j = 2\pi\sqrt{1/(2m\omega_{\bar{\ell}})}/\lambda_j$, where $\bar{\ell}$ denotes the propagation direction of the laser field $\Omega_j$ ($j=0,1$). With this correction, the driven-field Hamiltonian $H_\Omega$ becomes
\begin{equation}
    H'_\Omega= \sum_{j,\alpha}\frac{\Omega_j}{2} \left(e^{i\phi_j+i\eta_j(a_{\bar{\ell},\alpha}+{a_{\bar{\ell},\alpha}}^\dagger)}\ket{j}\bra{r_j}_\alpha+\text{h.c.}\right).
\end{equation}
For simplicity, in the expression of $\eta_j$ we use the wavelength $\lambda_j$ of the single-photon transition $\ket{j}\leftrightarrow\ket{r_j}$ (even if forbidden), which corresponds to a worst-case estimate; in practice, $\eta_j$ is significantly smaller when the Rydberg excitation involves a two- or three-photon transition.

\subsubsection{Unitary transformation}

To better illustrate the effects of the momentum kick, it is useful to perform the following unitary transformation
\begin{equation}
    \label{unitary_transf}
    U=\exp\left[-\sum_{\alpha,j}i\eta_{j}(a_{\bar{\ell},\alpha} + {a_{\bar{\ell},\alpha}}^\dag)\ket{r_j}\bra{r_j}_\alpha\right].
\end{equation}
It is straightforward to verify that the following transformation rules hold
\begin{align}
    U^\dag a_{\bar{\ell},\alpha}U & = a_{\bar{\ell},\alpha} +i \sum_{j=0,1} \eta_j\ket{r_j}\bra{r_j}_\alpha,\\
    U^\dag \ket{j}\bra{r_j}_\alpha U & = \exp\left[-i\eta_{j}(a_{\bar{\ell},\alpha} + {a_{\bar{\ell},\alpha}}^\dag)\right] \ket{j}\bra{r_j}_\alpha,\\
    U^\dag \ket{r_j}\bra{r_i}_\alpha U & = \exp\left[i(\eta_{j}-\eta_i)(a_{\bar{\ell},\alpha} + {a_{\bar{\ell},\alpha}}^\dag)\right] \ket{r_j}\bra{r_i}_\alpha.
\end{align}
Therefore, the kinetic, driving-laser-field, dipole–dipole interaction, and decay Hamiltonians transform, respectively, as
\begin{align}
    \label{kick_terms}
    U^\dag H_K U & = H_K + \frac{2\pi^2}{m\lambda^2}\sum_\alpha \mathbb{P}^r_\alpha +\frac{2\pi}{m\lambda}\sum_\alpha \mathbb{P}^r_\alpha p_\alpha^{\bar{\ell}},\\
    U^\dag H'_\Omega U & = \sum_{j,\alpha}\frac{\Omega_j}{2} \left(e^{i\phi_j}\ket{j}\bra{r_j}_{\alpha}+\text{h.c.}\right) = H_\Omega,\\
    U^\dag {H'}_J U & = H'_J,\\
    U^\dag {H}_\Gamma U & = H_\Gamma,
\end{align}
where it is assumed that $\eta_0 = \eta_1$, a good approximation, e.g., for $n = 100$ we have $|\eta_0 - \eta_1| \approx 10^{-6}$; here, $p^{\bar{\ell}}_\alpha$ denotes the momentum operator along the direction $\bar{\ell}$ of atom $\alpha$, and $\mathbb{P}^r_\alpha$ denotes the projector onto the Rydberg-state manifold of atom $\alpha$
\begin{equation}
    \mathbb{P}^r_\alpha = \sum_{j=0,1} \ket{r_j}\bra{r_j}_\alpha.
\end{equation}
Therefore, the only difference with respect to the unperturbed Hamiltonian is the presence of the two terms in Eq.~\eqref{kick_terms}, which encapsulate the twofold effect of photon recoil. The first term corresponds to a constant detuning of magnitude $2\pi^2/(m\lambda^2)$ on the Rydberg states and will be referred to as $H_{K}^{\text{detuning}}$, representing the kinetic energy imparted by a photon of the laser field with wavelength $\lambda$. The second term represents a coupling between the motional degrees of freedom along the $\bar{\ell}$ direction and the internal states, and will be referred to as $H_{K}^{\text{coupling}}$. This operator thus generates entanglement between the internal and motional states, resulting in decoherence of the former.

\subsection{\label{app:relevant_terms}Relevant terms}

Summarizing the results above, the total Hamiltonian $H'$, once corrected for the effects of distance fluctuations $H'_J$ in Eq.~\eqref{H'_J} and photon recoil, $H_K^{\text{detuning}}$ and $H_K^{\rm coupling}$ in Eq.~\eqref{kick_terms}, and including the vdW interactions $H_{\rm vdW}$ as well as the Rydberg decay non-Hermitian term $H_\Gamma$, becomes
\begin{align}
    \label{H'_complete}
    H' = H + H_{\rm vdW} +  H_{\rm vdW}^{(1,z)} + H_J^{(1,z)} + H_J^{(2,z)} + H_J^{(2,x)} + H_J^{(2,y)}+ H_{K}^{\text{detuning}} + H_{K}^{\text{coupling}}+H_\Gamma,
\end{align}
where $H$ is the unperturbed Hamiltonian, $H = H_\Omega + H_J + H_K$ (Eq.~1 in the main text). It is possible to estimate which of the nine noisy terms discussed above is most relevant, both numerically and analytically. In particular, in Tab.~\ref{tab:noisy_terms}, the coupling strength of each term is computed analytically, while the pulse infidelity is numerically simulated in the presence of only that term. This analysis is performed for both time-optimal and vdW-robust pulses. For the analytical estimation of the strength of each noisy term, the approach of Ref.~\cite{S_Bergonzoni_2025} is followed. The terms $H_J^{(1,z)}$ and $H_K^{\rm coupling}$ in Eq.~\eqref{H'_complete} are not resonant, i.e., the operators acting on the motional Hilbert space do not preserve the number of excitations, since the ladder operators are applied only once. In contrast quadratic terms such as $H_J^{(2,z)}$ contain combinations of ladder operators of the form $a_{z,A}a_{z,B}^\dag$ that preserve the total energy. However, given that the motional detuning $\omega_z/2\pi = 100\,\mathrm{kHz}$ is small compared to the interaction energy scale $J/2\pi \approx 1\text{–}10\,\mathrm{MHz}$, the off-resonant terms can be treated analogously to the resonant ones. For example, the magnitude of the first-order term in the distance fluctuations, $H_J^{(1,z)}$, can be evaluated as
\begin{equation}
    \label{Delta_E_J}
    \frac{\Delta E^{(1,z)}_J}{J} = \frac{3 \sigma_{z_A-z_B}}{R} = \frac{3}{R} \sqrt{\frac{1}{2m\omega_z} \coth\left(\frac{\omega_z}{2k_BT_{\text{temp}}}\right)},
\end{equation}
with $\sigma_{z_A-z_B}$ the standard deviation of the displacement operator $z_A-z_B$, evaluated over a thermal distribution of the states. Similarly, the magnitude of the other motion-related terms are
\begin{align}
    \frac{\Delta E^{(2,z)}_J}{J} & = \frac{6 \sigma_{(z_A-z_B)^2}}{R^2} = \frac{6\sqrt{2}}{m\omega_z R}  \coth\left(\frac{\omega_z}{2k_BT_{\text{temp}}}\right),\\
    \frac{\Delta E^{(2,x)}_J}{J} & = \frac{3 \sigma_{(x_A-x_B)^2}}{R^2} = \frac{3\sqrt{2}}{m\omega_x R}  \coth\left(\frac{\omega_x}{2k_BT_{\text{temp}}}\right),\\
    \frac{\Delta E^{(2,y)}_J}{J} & = \frac{3 \sigma_{(y_A-y_B)^2}}{R^2} = \frac{3\sqrt{2}}{m\omega_y R}  \coth\left(\frac{\omega_y}{2k_BT_{\text{temp}}}\right),\\
    \frac{\Delta E^{(1,z)}_{\rm vdW}}{V_{\rm vdW}} & = \frac{6 \sigma_{z_A-z_B}}{R} = \frac{6}{R} \sqrt{\frac{1}{2m\omega_z} \coth\left(\frac{\omega_z}{2k_BT_{\text{temp}}}\right)},  
\end{align}
where $\sigma_{(z_A-z_B)^2}=\sqrt{2}\sigma_{z_A-z_B}^2$ is the standard deviation of the squared of the displacement operator $(z_A-z_B)^2$. The magnitude of the motion-internal space coupling emerging from photon recoil, $H_K^{\text{coupling}}$, can be similarly evaluated as
\begin{equation}
    \frac{\Delta E_K^{\text{coupling}}}{2\pi} = \frac{1}{\lambda} \sqrt{\frac{\omega_{\bar{\ell}}}{m} \coth \left(\frac{\omega_{\bar{\ell}}}{2k_B T_{\text{temp}}}\right)}.
\end{equation}
Table~\ref{tab:noisy_terms} shows, consistently with both the analytical estimates of the coupling strengths and the numerical fidelity calculations, that the main source of error affecting the time-optimal pulses arises from the vdW interactions. Once these forces are included in the optimization and vdW–robust pulses are obtained, the dominant error sources become, in decreasing order, the Rydberg decay $H_\Gamma$, the first-order atomic displacement along the interatomic $z$ direction $H_J^{(1,z)}$, and the photon-recoil coupling term $H_K^{\rm coupling}$. Note that the coupling term associated with spontaneous emission, reported in Tab.~\ref{tab:noisy_terms} as the decay rate $\Gamma$, is smaller than the others; however, it represents the magnitude of a non-Hermitian contribution to the Hamiltonian and is therefore not directly comparable to the other (Hermitian) noise terms.

\begin{table*}[t]
\centering
\renewcommand{\arraystretch}{1.2}
\begin{tabular}{cc|c|c|c}
\multirow{2}{*}{\textbf{Noise term}} & \multirow{2}{*}{} &
\multicolumn{2}{c|}{\centering$\mathbf{1-F}$} &
\multirow{2}{*}{\textbf{Coupling strength ($2\pi\times$Hz)}} \\
\cline{3-4}
 &  & \textbf{Time-optimal}  & \textbf{vdW-robust}  &   \\
\hline
 No noise & $\varnothing$ & $7.0\times10^{-6}$ & - & $0$ \\
\hline
 vdW interactions & $H_V$ & $1.3\times10^{-2}$ & $2.0\times10^{-4}$ & $9.6\times10^5$ \\
\hline
 Rydberg decay & $H_\Gamma$ & $7.0\times10^{-4}$ & $7.0\times10^{-4}$ & $1.3\times10^3$ \\
\hline
 1st order motion along $z$ ($J$ term) & $H_J^{(1,z)}$ & $4.3\times10^{-5}$ & $3.7\times10^{-4}$ & $2.4 \times10^4$ \\
\hline
 2nd order motion along $x$ & $H_J^{(2,x)}$ & $<10^{-7}$ & $<10^{-7}$ & $6.6\times10^2$ \\
\hline
2nd order motion along $y$ & $H_J^{(2,y)}$ & $<10^{-7}$ & $<10^{-7}$ & $5.9\times10$ \\
\hline
2nd order motion along $z$ & $H_J^{(2,z)}$ & $<10^{-7}$ & $<10^{-7}$ & $1.2\times10^2$ \\
\hline
 1st order motion along $z$ ($V_{\rm vdW}$ term) & $H_{\rm vdW}^{(1,z)}$ & $<10^{-7}$ & $<10^{-7}$ & $9.2\times10^3$ \\
\hline
\multirow{2}{*}{Photon recoil  $H_K^{\text{detuning}}$ with} & $\Delta_j =0$ & $1.3\times10^{-5}$ & $8.7\times10^{-6}$ & \multirow{2}{*}{$2.6\times10^4$} \\ 
 & $\Delta_j = 2\pi^2/m\lambda^2$ & $<10^{-7}$ & $<10^{-7}$  & \\
\hline
\multirow{2}{*}{Photon recoil $H_K^{\text{coupling}}$ with} 
 & $\bar{\ell}=x$ & $2.3\times10^{-5}$ & $1.9\times10^{-5}$  & $7.3\times10^4$ \\ 
 & $\bar{\ell}=z$ & $4.9\times10^{-5}$ & $5.1\times10^{-5}$  & $4.9\times10^4$\\ 
 \hline
 Scattering in the intermediate $\ket{e}$ state & $\Delta_{2p}=40\Omega_{\rm max}$ & $3.5\times10^{-3}$ & $2.9\times{10^{-3}}$ & - \\
\hline
Off-resonant Rydberg transitions & $B=50\,{\rm G}$ & $2.0\times{10^{-3}}$ & $1.2\times10^{-3}$ & - \\
\hline
All &  & $1.9\times10^{-2}$ & $5.1\times10^{-3}$ & - \\
\end{tabular}
\caption{To compute these terms, the following parameters were adopted: a pulse with $\Omega_{\text{max}}/J = 2.1$, temperature $T_{\text{temp}} = 1\,\mathrm{\mu K}$, maximum Rabi frequency $\Omega_{\text{max}}/2\pi = 10\,\mathrm{MHz}$, trapping frequencies $\omega_z/2\pi = 100\,\mathrm{kHz} = 5\omega_x/2\pi$, and Rydberg states with principal quantum number $n = 100$. The value $1-F = 7.0 \times 10^{-6}$ (first line) represents the numerical zero at which GRAPE optimization is stopped for the time-optimal pulse. For the vdW–robust pulse, the optimization is halted at $1-F = 2.0 \times 10^{-4}$ to avoid an excessive gate duration and increased decay errors. The coupling strengths are computed from the prefactors in the analytical expression of the Hamiltonian $H'$, with $\hbar = 1$.}
\label{tab:noisy_terms}
\end{table*}

\subsection{Recapture probability}
Once the excitation pulse is applied and the optical tweezers are turned back on, the scheme is expected to ensure a high recapture probability of the atoms \cite{S_deKeijzer_2023, S_Emperauger_2025}. The dipolar force exerted by one atom on the other is given by $F = |\partial J / \partial R| = 3C_3 / R^4$, corresponding to an acceleration $a = 3C_3 / (m R^4)$ on each atom. Since the interatomic distance $R$ is large compared to the scale of atomic motion, the acceleration can be considered constant during the pulse sequence. The resulting displacement of an atom during the pulse can thus be estimated as
\begin{equation}
\Delta z = \frac{1}{2}aT^2 = \frac{3}{2}\frac{C_3 T^2}{m R^4} \approx 0.06\,\mathrm{nm},
\end{equation}
using the parameters $n = 100$, $T = 190\,\mathrm{ns}$, $R = 19.7\,\mathrm{\mu m}$, and $m = 1.4 \times 10^{-25}\,\mathrm{kg}$ (vdW-robust pulse for $\Omega_{\rm max}/J = 2.1$). This displacement is several orders of magnitude smaller than the characteristic harmonic oscillator length, $z_{\text{osc}} = \sqrt{1/(2 m \omega_z)} \approx 24\,\mathrm{nm}$ for $\omega_z/2\pi = 100\,\mathrm{kHz}$, and therefore the recapture of the atoms should be straightforward.

\section{\label{app:fidelity}Gate fidelity with mixed states}

In the cost functional employed by GRAPE to optimize both the time-optimal and vdW–robust pulses implementing the iSWAP gate, the Bell-state fidelity $F$ is used as the figure of merit
\begin{equation}
    \label{F_supp}
    F = \frac{1}{16}\left| \sum_{q} \braket{\psi_q| R^{\otimes2}U_{\rm{iSWAP}}|q} \right|^2,
\end{equation}
where $\ket{\psi_q}$ is the actual two-qubit final state obtained by applying the pulse to $\ket{q}$, from the computational basis $\ket{q} \in \{\ket{00}, \ket{01}, \ket{10}, \ket{11}\}$. $\ket{\psi_q}$ is therefore compared to the target state $U_{\rm{iSWAP}}\ket{q}$ up to a final global single-qubit operation $R(\theta,\varphi,\lambda) = e^{i(\varphi+\lambda)/2} R_z(\varphi) R_y(\theta) R_z(\lambda)$, where $\varphi$, $\theta$, and $\lambda$ are rotation angles treated as additional optimization parameters.
The expression in Eq.~\eqref{F_supp} is convenient because it guarantees not only the correct population transfer among the computational basis states but also the preservation of coherence, i.e., the relative phases between them. However, this form cannot be directly applied to evaluate the fidelity of the pulses in the presence of atomic motion. When coupling to the motional degrees of freedom is included, the final internal state generally becomes entangled with the motional state and is thus described by a mixed density matrix rather than a pure state. Furthermore, the fidelity then depends on the initial motional state, requiring an average over all possible initial conditions. To properly account for these effects, we adopt the following fidelity function
\begin{equation}
    \label{F_complete}
    F' = \sum_{m=0}^M \frac{b_m}{\dim{Q}^2} \left| \sum_{q} \sqrt{\mathcal{F}(\rho_{qm},\rho_q^{\text{targ}})} e^{i\phi_{qm}} \right|^2.
\end{equation}
The first sum in Eq.~\eqref{F_complete} runs over all harmonic oscillator eigenstates $\ket{m}$ with eigenenergy $E_m$, where $b_m = \exp(-E_m/k_B T_{\text{temp}})/\mathcal{Z}$ is the corresponding Boltzmann factor at temperature $T_{\text{temp}}$, $k_B$ is the Boltzmann constant, and $\mathcal{Z} = \sum_{m=0}^M \exp(-E_m/k_B T_{\text{temp}})$ is the partition function. Accordingly, we assume a thermal distribution of the initial motional excitations. The cutoff $M$ must be chosen sufficiently large to ensure an accurate description of the system.
The second sum runs over all computational basis states $\ket{q}$. The final reduced density matrix $\rho_{qm}$ is obtained by evolving the pure product state $\ket{q,m}$ under the gate unitary $U(T,0)$ and tracing out the motional degrees of freedom,
$\rho_{qm} = \Tr_{\text{mot}}\{ U(T,0)\ket{q,m}\bra{q,m}U(T,0)^\dag \}$.
The target density matrix $\rho_q^{\text{targ}}$ is instead obtained by applying the ideal gate $U_{\rm iSWAP}$ to the internal state $\ket{q}$ in the absence of motion and other noise sources. Finally, $\mathcal{F}(\rho,\sigma)$ denotes the Uhlmann fidelity, which quantifies the overlap between two density matrices \cite{S_Jozsa_1994, S_Nielsen_Chuang_2010}.
\begin{equation}
    \mathcal{F}(\rho,\sigma) = \left(\Tr\sqrt{\sqrt{\rho}\sigma \sqrt{\rho}}\right)^2.
\end{equation}
The Uhlmann fidelity is a real quantity that quantifies the overlap between the target pure state $\rho_q^{\text{targ}}$ and the actual reduced density matrix $\rho_{qm}$ \cite{S_Uhlmann_1976}. However, it does not account for the coherence between different computational basis states $q$. To incorporate this information, we multiply the Uhlmann fidelity by the accumulated phase $\phi_{qm} = \arg(\braket{\psi^{\text{targ}}_q | \psi_{qm}})$, where $\ket{\psi_{qm}}$ denotes the actual final state (including both internal and motional degrees of freedom) evolved from the initial state $\ket{q,m}$, and $\ket{\psi^{\text{targ}}_q} = U_\text{iSWAP}\ket{q}$ represents the ideal target state for $\ket{q}$, formally tensored with a uniform superposition (all-ones vector) in the motional subspace.\\
The fidelity function in Eq.~\eqref{F_complete} ensures that, for every motional state $m<M$ and every internal state $q$ in the computational basis, the pulse drives the system to the target state $\ket{\psi_q^{\text{targ}}}$ up to a phase factor $\phi_{qm}$, which must be identical for all $q$ in order to preserve coherence.

\section{Performance with noise }

\begin{figure*}
    \centering
    \includegraphics[width=0.93\linewidth]{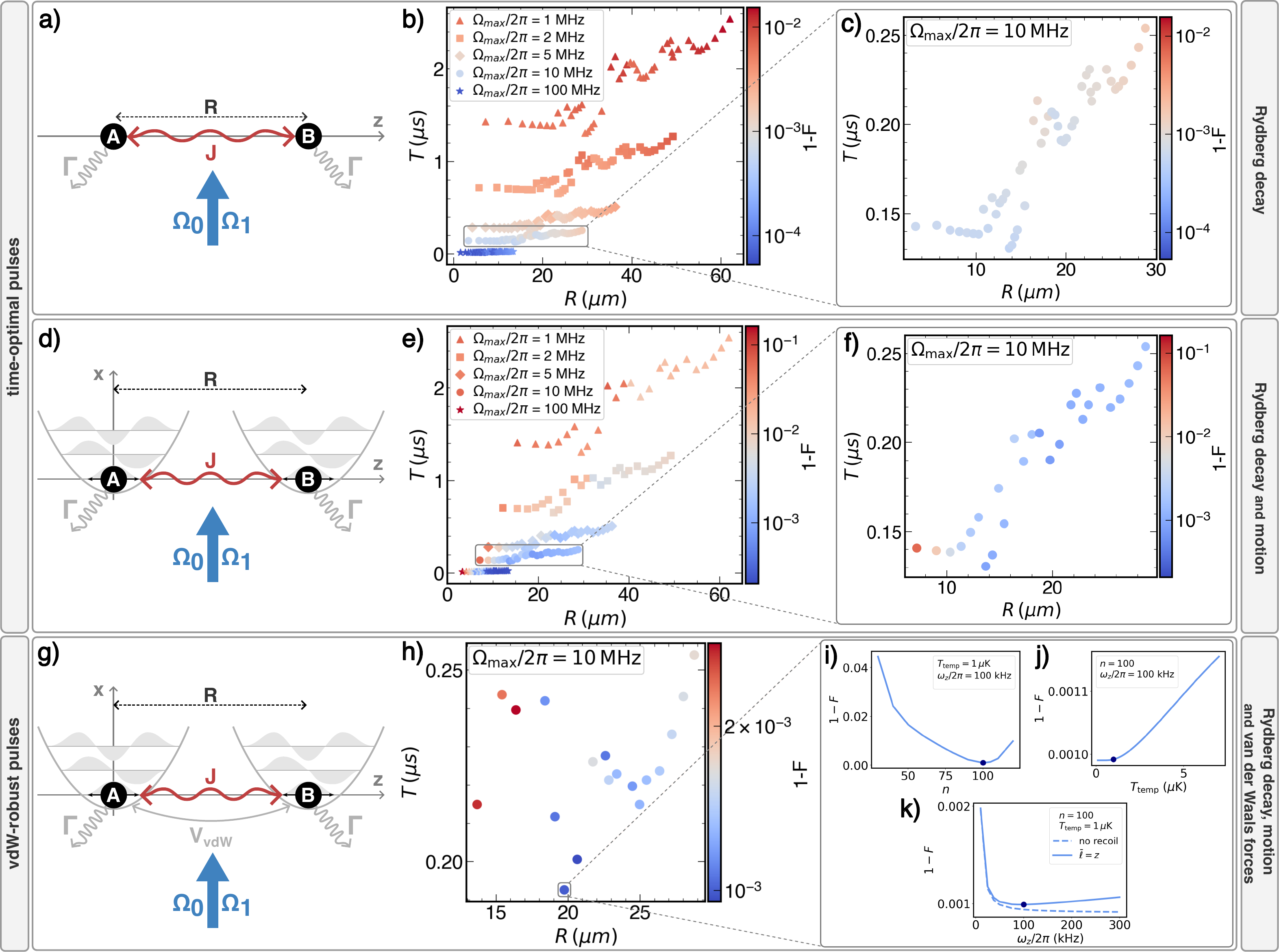}
    \caption{\textbf{(a)} Schematic representation of the two interacting atoms at a distance $R$, with global external driving fields $\Omega_j$ ($j=0,1)$, and with decay rate $\Gamma_j$. \textbf{(b)} Fidelity of the time-optimal pulses [see Fig.~\ref{fig:S_pulses}] in the presence of spontaneous emission, with $n=100$ and different Rabi frequencies $\Omega_{\rm max}$. The time duration $T$ and the range $R$ of each pulse depends on the ratio $\Omega_{\text{max}}/J$ and the maximum Rabi frequency $\Omega_{\text{max}}$. \textbf{(c)} Zoom-in on the data points for $\Omega_{\text{max}}/2\pi=10\,\mathrm{MHz}$. \textbf{(d)} Schematic representation of the two interacting atoms at a distance $R$, with global external driving fields $\Omega_j$ ($j=0,1)$, with decay rate $\Gamma_j$ and moving inside the radial harmonic-like trapping potential of the tweezers. \textbf{(e)} Fidelity of the time-optimal pulses [see Fig.~\ref{fig:S_pulses}] in the presence of spontaneous emission and radial motion, with $n=100$ and different Rabi frequencies $\Omega_{\rm max}$. The time duration $T$ and the range $R$ of each pulse depends on the ratio $\Omega_{\text{max}}/J$ and the maximum Rabi frequency $\Omega_{\text{max}}$. \textbf{(f)} Zoom-in on the data points for $\Omega_{\text{max}}/2\pi=10\,\mathrm{MHz}$. \textbf{(g)} Schematic representation of the two interacting atoms at a distance $R$, with global external driving fields $\Omega_j$ ($j=0,1)$, with decay rate $\Gamma_j$, moving inside the radial harmonic-like trapping potential of the tweezers, and with van der Waals interactions $V_{\rm vdW}$.  \textbf{(h)} Fidelity of the vdW-robust pulses in the presence of spontaneous emission, radial motion and vdW interactions, with $n=100$ and $\Omega_{\text{max}}/2\pi=10\,\mathrm{MHz}$. The time duration $T$ and the range $R$ of each pulse depends on the ratio $\Omega_{\text{max}}/J$ and the maximum Rabi frequency $\Omega_{\text{max}}$. \textbf{(i)} Fidelity $F$ of the time-optimal pulse for $\Omega_{\text{max}}/J=2.1$ as a function of the principal quantum number $n$. \textbf{(j)} Fidelity $F$ of the time-optimal pulse for $\Omega_{\text{max}}/J=2.1$ as a function of the temperature $T_{\text{temp}}$. \textbf{(k)} Fidelity $F$ of the time-optimal pulse for $\Omega_{\text{max}}/J=2.1$ as a function of the radial trapping frequency $\omega_{z}/2\pi$. The solid line corresponds to the case where the momentum kick occurs along the interatomic direction ($\bar{\ell}=z$). The dashed line corresponds to the case where there is no photon recoil}
    \label{fig:S_fidelity}
\end{figure*}

In this section, the performance of the time-optimal and vdW-robust pulses is evaluated in the presence of different error sources, namely the finite lifetime of the Rydberg states, the coupling to motional degrees of freedom, and the residual off-resonant vdW interactions. All results are summarized in Fig.~\ref{fig:S_fidelity}.

\subsection{Performance with spontaneous emission}

In Figs.~\hyperref[fig:S_pulses]{\ref{fig:S_pulses}(b)-(c)}, the fidelity of the time-optimal pulses is analyzed, including the decay term $H_\Gamma$. One can observe that, for a fixed interatomic distance $R$, larger Rabi frequencies $\Omega_{\text{max}}$ lead to higher fidelities, as they make the pulses faster and reduce the time spent in the Rydberg states. The same trend holds for shorter distances $R$, since the interaction strength $J \sim R^{-3}$ implies that the gate duration $T$ is generally shorter for small $R$, resulting in less time spent in the Rydberg states and reduced susceptibility to decay. There are some values of the ratio $\Omega_{\text{max}}/J$ — thus of the distance $R$ — for which the duration $T$ is particularly short with respect to the nearby pulses, e.g., $T\Omega_{\text{max}}=0.7$ [see Fig.~\hyperref[fig:S_pulses]{\ref{fig:S_pulses}(c)}] and $T\Omega_{\text{max}}=2.1$ [see Fig.~\hyperref[fig:S_pulses]{\ref{fig:S_pulses}(d.2)}], therefore these points realize particularly good pulses. For example, for $\Omega_{\text{max}}/J=2.1$, the time spent in Rydberg states is $T_{\text{Ryd}}\Omega_{\text{max}}=10.6$, and choosing $\Omega_{\text{max}}=10\,\mathrm{MHz}$ results in an infidelity $1-F=7.0\times10^{-4}$.

\subsection{\label{sec:performance_motion}Performance with motion}

In Figs.~\hyperref[fig:S_fidelity]{\ref{fig:S_fidelity}(e)–(f)}, the fidelity of the optimal pulses is analyzed, taking into account both the finite lifetime of the Rydberg states and the atomic motion along the radial $z$ direction (including distance fluctuations and photon recoil). As discussed above, for a fixed interatomic distance $R$, increasing the maximum Rabi frequency $\Omega_{\text{max}}$ generally improves the fidelity, as the gate becomes faster. However, in contrast to the case where only decay is considered, for a fixed Rabi frequency $\Omega_{\text{max}}$, the shortest pulses are not necessarily those with the highest fidelity [see Fig.~\hyperref[fig:S_pulses]{\ref{fig:S_pulses}(f)}]. This occurs because, at short distances, the effect of atomic motion is much stronger and dominates over spontaneous emission — indeed, Eq.~\eqref{Delta_E_J} shows that the former scales as $1/R$. Conversely, for very long-range gates, where the effects of atomic motion are no longer significant, the fidelity decreases due to the finite lifetime of the Rydberg states and the longer pulse durations. Optimal performance is therefore achieved at intermediate distances. For example, the pulse with $\Omega_{\text{max}}/J=2.1$ [see Fig.~\hyperref[fig:S_pulses]{\ref{fig:S_pulses}(d.2)}] achieves an infidelity of $1-F=7.8\times10^{-4}$ for $\Omega_{\text{max}}/2\pi=10\,\mathrm{MHz}$, with $T_{\text{temp}}=1\,\mathrm{\mu K}$.

\subsection{\label{sec:performance_vdW}Performance with van der Waals interactions}

 Through numerical simulations, we verified that the vdW interactions do not significantly affect the population dynamics during the time-optimal pulses, but they lead to the accumulation of an unwanted phase, particularly on the qubit state $\ket{00}$, since $V_{00}$ is the largest vdW interaction strength (see Tab.~\ref{tab:atomic_data1}). The time-optimal pulse with $\Omega_{\text{max}}/J = 2.1$, using $n = 100$ and $\Omega_{\text{max}}/2\pi = 10\,\mathrm{MHz}$, and including the vdW interactions $H'_{\rm vdW}$, yields an infidelity of $1-F = 1.3 \times 10^{-2}$ — an order of magnitude larger than in the case with only decay and motion. This infidelity can be reduced in two ways: \textit{(i)} Keep the same pulses shown in Fig.~\ref{fig:S_pulses} and determine the optimal principal quantum number $n$ that maximizes $F$, balancing longer Rydberg lifetimes (for large $n$) against weaker vdW interactions (for small $n$). For example, for the time-optimal pulse with $\Omega_{\text{max}}/J=2.1$, the optimal principal quantum number is $n\approx70$, with an infidelity $1-F=3.4\times10^{-3}$, but also a reduced range. \textit{(ii)} Fix the highest achievable principal quantum number (to minimize decay errors), e.g., $n = 100$, and find new pulses that are inherently robust to vdW interactions by including the vdW Hamiltonian $H_{\rm vdW}$ directly in the GRAPE optimization. The time-optimal pulses are used as initial guesses, and, if necessary, their duration $T$ can be slightly increased to adapt to the presence of vdW interactions \cite{S_Giudici_2025}. For each value of $\Omega_{\text{max}}/J$, there exists a different vdW–robust pulse corresponding to each ratio $\Omega_{\text{max}}/V_{00}$; thus, the robust pulses also depend explicitly on the value of $\Omega_{\text{max}}$. This second strategy consistently produces higher fidelities, and we therefore adopt it. In general, for small ratios $\Omega_{\text{max}}/J < 1$, GRAPE fails to find robust pulses with durations comparable to the time-optimal ones [see Fig.~\hyperref[fig:S_fidelity]{\ref{fig:S_fidelity}(h)}]. This is because small $\Omega_{\text{max}}/J$ also implies small $R$, which enhances the effect of vdW interactions. Conversely, for large ratios $\Omega_{\text{max}}/J \gtrsim 3$, GRAPE can reduce the sensitivity to vdW interactions while keeping the same pulse duration. Intermediate ratios $\Omega_{\text{max}}/J$ yield robust pulses at the cost of slightly longer durations. In Fig.~\hyperref[fig:S_fidelity]{\ref{fig:S_fidelity}(h)}, the performance of a group of vdW-robust pulses for different values of the ratio $\Omega_{\text{max}}/J$ is analyzed. In Figs.~\hyperref[fig:S_fidelity]{\ref{fig:S_fidelity}(e)-(g)}, the fidelity of the robust pulse $\Omega_{\text{max}}/J=2.1$ is analyzed as a function of the principal quantum number $n$, the temperature $T_{\rm temp}$ and the radial trapping frequency $\omega_z$. For $n=100$, $T_{\rm temp}=1\,{\rm \mu K}$ and $\omega_z=100\,{\rm kHz}$ the total infidelity, including all the previous physical effects is $1-F=9.9\times10^{-4}$.

\subsection{\label{sec:performance_transitions} Performance with realistic atomic transitions}

In order to evaluate the gate fidelity while accounting for the complexity of realistic atomic transitions, we include in our simulations both multi-photon transitions and off-resonant Rydberg states. Each atom is modeled as a seven-level system $\{\ket{0}, \ket{1}, \ket{e}, \ket{r_0'}, \ket{r_0}, \ket{r_1'}, \ket{r_1}\}$, where $\ket{e}$ is the intermediate state for the two-photon transition $\ket{0}\leftrightarrow\ket{r_0}$, as shown in Fig.~\hyperref[fig:atomic_transitions]{\ref{fig:atomic_transitions}(a)}, and $\ket{r_j'}$ ($j=0,1$) are the two dominant off-resonant Rydberg states that interfere with the excitation to the target states $\ket{r_j}$, as shown in Fig.~\hyperref[fig:atomic_transitions]{\ref{fig:atomic_transitions}(c)}.\\
A crucial factor in determining the final fidelity is the intermediate detuning $\Delta_e$ from the intermediate state $\ket{e}$, as well as the magnetic field $B$. The detuning $\Delta_e$ controls the population of $\ket{e}$ and, consequently, the susceptibility of the gate to spontaneous emission: the larger $\Delta_e$, the shorter the time spent in $\ket{e}$. However, larger values of $\Delta_e$ also require stronger laser power to maintain the same effective Rabi frequency $\Omega^{\mathrm{eff}}_0$. In the simulations, spontaneous emission from the intermediate state is modeled via a non-Hermitian term in the Hamiltonian, which likely results in an overestimate of the infidelity.\\
The magnetic field, in turn, plays a crucial role by determining the detuning of the off-resonant states $\ket{r_j'}$ and, therefore, how strongly they influence the gate dynamics, either through population leakage or by inducing Stark shifts on the other states. In particular, the states $\ket{r_1}$ and $\ket{r_1'}$ are degenerate in the absence of a magnetic field.\\
The vdW-robust pulse for $\Omega_{\rm max}/J = 2.1$, with $\Omega_{\rm max} = 10\,\mathrm{MHz}$ and $n = 100$, has been simulated including these two error sources, with $B = 50\,\mathrm{G}$ and $\Delta_e = 40\,\Omega_{\rm max}$. This yields an infidelity of $1 - F = 4.1 \times 10^{-3}$, almost equally distributed between scattering from the state $\ket{e}$ and off-resonant couplings, as shown in Tab.~\ref{tab:noisy_terms}. The overall infidelity, including also the previously discussed effects, can be (over)estimated by summing this contribution with that arising from the other error sources, leading to $1 - F \approx 5.1 \times 10^{-3}$. We emphasize that, even when using more conservative estimates — such as a smaller Rabi frequency $\Omega_{\rm max} = 5\,\mathrm{MHz}$ and the vdW-robust pulse corresponding to $\Omega_{\rm max}/J = 4.25$, with $n = 100$, $B = 20\,\mathrm{G}$, and $\Delta_e = 40\,\Omega_{\rm max}$ — the overall infidelity, accounting for both realistic multi-photon transitions and the other error sources, is $1 - F \approx 8.4 \times 10^{-3}$. This value remains below the threshold of most quantum error-correction schemes and, at the same time, provides a viable route to entangle atoms separated by $R = 31\,\mu\mathrm{m}$ within a gate time of $T = 430\,\mathrm{ns}$.

\end{document}